\documentclass[12pt]{article}
\pdfoutput=1
\usepackage{a4wide,epsfig,psfrag,amsmath,amssymb,cite,scalefnt}
\usepackage{color}
\usepackage{amsmath,comment,braket}
\usepackage{placeins}

\graphicspath{ {figs_tot/} }

\newcommand{\vph}{\vphantom{\Big\{}}

\newcommand{\lms}{l_{ms}}
\newcommand{\lts}{l_{ts}}
\newcommand{\lots}{l_{1ts}}

\newcommand{\be}{\begin{equation}}
\newcommand{\ee}{\end{equation}}
\newcommand{\bea}{\begin{eqnarray}}
\newcommand{\eea}{\end{eqnarray}}

\allowdisplaybreaks

\newcommand{\gsim}{\;\rlap{\lower 3.5 pt \hbox{$\mathchar \sim$}} \raise 1pt
 \hbox {$>$}\;}
\newcommand{\lsim}{\;\rlap{\lower 3.5 pt \hbox{$\mathchar \sim$}} \raise 1pt
 \hbox {$<$}\;}

\begin{document}

\title{\vskip-3cm{\baselineskip14pt
    \begin{flushleft}
      \normalsize TTP18-039
  \end{flushleft}}
  \vskip1.5cm
  Double Higgs boson production at NLO in the high-energy limit: complete
  analytic results
}

\author{
  Joshua Davies$^{a}$,
  Go Mishima$^{a,b}$,
  Matthias Steinhauser$^{a}$,
  David Wellmann$^{a}$
  \\[1mm]
  {\small\it $^a$Institut f{\"u}r Theoretische Teilchenphysik}\\
  {\small\it Karlsruhe Institute of Technology (KIT)}\\
  {\small\it Wolfgang-Gaede Stra\ss{}e 1, 76128 Karlsruhe, Germany}
  \\[1mm]
  {\small\it $^b$Institut f{\"u}r Kernphysik}\\
  {\small\it Karlsruhe Institute of Technology (KIT)}
  \\
  {\small\it Hermann-von-Helmholtz-Platz 1, 76344 Eggenstein-Leopoldshafen, Germany}
}
  
\date{}

\maketitle

\thispagestyle{empty}

\begin{abstract}

  We compute the NLO virtual corrections to the partonic cross section of
  $gg\to HH$, in the high energy limit.  Finite Higgs boson mass effects are
  taken into account via an expansion which is shown to converge quickly.
  We obtain analytic results for the next-to-leading order form factors
  which can be used to compute the cross section.  The method used for the
  calculation of the (non-planar) master integrals is described in
  detail and explicit results are presented.
  
%

\end{abstract}

\thispagestyle{empty}

\sloppy


\newpage


\section{\label{sec::intro}Introduction}

Higgs boson pair production is a promising channel to investigate the
self interaction of the Higgs boson.  Although it is very challenging from the
experimental point of view it is expected that after the high-luminosity
upgrade of the LHC constraints on the Higgs boson tri-linear coupling will be able
to be obtained. In order to determine whether or not the Higgs sector is Standard
Model--like it is therefore important to have the higher order corrections to
double Higgs boson production under control. A further building block towards
this goal is considered in this paper by providing analytic results in the
high-energy limit.

Higgs boson pairs are predominantly produced by the gluon-fusion channel
and in the recent years a number of higher order corrections have been
computed to $gg\to HH$, both for the total cross section and for differential
distributions. We refrain from providing a detailed review but refer to
Ref.~\cite{Grazzini:2018bsd} where several recent results are combined and 
approximate next-to-next-to-leading order (NNLO) expressions are constructed.

From the technical side the main new ingredients from this paper are analytic
results for the two-loop non-planar master integrals for $gg\to HH$
which, in combination with the findings of Ref.~\cite{Davies:2018ood},
allows one to obtain the next-to-leading order (NLO) amplitude for this process in
the high-energy limit. This complements the NLO results obtained from the large
top quark--mass {expansion~\cite{Dawson:1998py,Grigo:2014jma,Degrassi:2016vss}},
from the threshold expansion~\cite{Grober:2017uho} and from an expansion 
for small Higgs transverse momentum~\cite{Bonciani:2018omm}.
Furthermore, it provides an important cross check and eventually an
alternative approach to the exact result obtained in
{Refs.~\cite{Borowka:2016ehy,Borowka:2016ypz,Baglio:2018lrj}} using a numerical approach.
Recently it has been suggested to expand the $gg\to HH$ amplitude
only in the Higgs boson mass but keep the dependence on the kinematic
invariants and the top quark mass~\cite{Xu:2018eos}. This also leads to
simpler expressions, however, one still has to solve integrals involving three
scales.

To describe the amplitude $g(q_1)g(q_2)\to H(q_3)H(q_4)$, with all
momenta $q_i$ defined to be incoming, we introduce the Mandelstam variables as follows
\begin{eqnarray}
  \tilde{s}=(q_1+q_2)^2\,,\qquad \tilde{t}=(q_1+q_3)^2\,,\qquad \tilde{u}=(q_2+q_3)^2\,,
  \label{eq::stu}
\end{eqnarray}
with
\begin{eqnarray}
  q_1^2=q_2^2=0\,,\qquad  q_3^2=q_4^2=m_H^{2}\,,\qquad
  \tilde{s}+\tilde{t}+\tilde{u}=2m_H^{2}\,.
  \label{eq::q_i^2}
\end{eqnarray}
As described in more detail
in Subsection~\ref{sub::expMh} we perform an expansion in the Higgs
boson mass. This means that we use the kinematics defined in
Eqs.~(\ref{eq::stu}) and~(\ref{eq::q_i^2}) when evaluating the
amplitude, but before evaluating the Feynman integrals we set $m_H=0$ and
obtain the following variables which are relevant for the computation of the
integrals\footnote{In the limit $m_H=0$ we drop the tilde from the Mandelstam variables.}
\begin{eqnarray}
  s=2q_1\cdot q_2\,,\qquad t=2q_1\cdot q_3\,,\qquad u=2q_2\cdot
  q_3=-s-t
  \,.
\end{eqnarray}
Thus the integrals will only depend on the variables $s,t$ and $m_t^{2}$, and
when computing them we further assume that $m_t^2\ll s,|t|$. It is convenient to
introduce the scattering angle $\theta$ of the Higgs boson in the
center-of-mass frame which leads to the following relation in terms of these
variables,
\begin{eqnarray}
\label{eqn:thetadef}
  t &=& -\frac{s}{2}\left(1-\cos{{\theta}}\right)
  \,.
\end{eqnarray}

Due to Lorentz and gauge invariance it is possible to define two scalar matrix
elements ${\cal M}_1$ and ${\cal M}_2$ as
\begin{eqnarray}
  {\cal M}^{ab} &=& 
  \varepsilon_{1,\mu}\varepsilon_{2,\nu}
  {\cal M}^{\mu\nu,ab}
  \,\,=\,\,
  \varepsilon_{1,\mu}\varepsilon_{2,\nu}
  \delta^{ab}
  \left( {\cal M}_1 A_1^{\mu\nu} + {\cal M}_2 A_2^{\mu\nu} \right)
  \,,
\end{eqnarray}
where $a$ and $b$ are adjoint colour indices and the two Lorentz structures
are given by
\begin{eqnarray}
  A_1^{\mu\nu} &=& g^{\mu\nu} - {\frac{1}{q_{12}}q_1^\nu q_2^\mu
  }\,,\nonumber\\
  A_2^{\mu\nu} &=& g^{\mu\nu}
                   + \frac{1}{q_T^2 q_{12}}\left(
                   q_{33}    q_1^\nu q_2^\mu
                   - 2q_{23} q_1^\nu q_3^\mu
                   - 2q_{13} q_3^\nu q_2^\mu
                   + 2q_{12} q_3^\mu q_3^\nu \right)\,,
\end{eqnarray}
with
\begin{eqnarray}
  q_{ij} &=& q_i\cdot q_j\,,\qquad
  q_T^{\:2} \:\:\:=\:\:\: \frac{2q_{13}q_{23}}{q_{12}}-q_{33}
  \,.
\end{eqnarray}
The Feynman diagrams involving the triple Higgs boson coupling only
contribute to $A_1^{\mu\nu}$ and, thus, it is convenient to decompose
${\cal M}_1$ and ${\cal M}_2$ into ``triangle'' and ``box'' form factors
\begin{eqnarray}
  {\cal M}_1 &=& X_0 \, s \, \left(\frac{3 m_H^2}{s-m_H^2} F_{\rm tri} + F_{\rm box1}\right)
                 \,,\nonumber\\
  {\cal M}_2 &=& X_0 \, s \, F_{\rm box2}
                 \,,
                 \label{eq::calM}
\end{eqnarray}
with
\begin{eqnarray}
  X_0 &=& \frac{G_F}{\sqrt{2}} \frac{\alpha_s(\mu)}{2\pi} T \,,
\end{eqnarray}
where $T=1/2$ and $\mu$ is the renormalization scale.  We furthermore define
the expansion in $\alpha_s$ of the form factors as
\begin{eqnarray}
  F &=& F^{(0)} + \frac{\alpha_s(\mu)}{\pi} F^{(1)} + \cdots
  \,,
  \label{eq::F}
\end{eqnarray}
and similarly for ${\cal M}_i$.  Throughout this paper the strong
coupling constant is defined with six active quark flavours.  Note that the
form factors are defined such that the one-loop colour factor $T$ is contained in
the prefactor $X_0$.

The main results of this paper can be summarized as follows:
\begin{itemize}
\item We compute all planar (see Ref.~\cite{Davies:2018ood}) and non-planar master
  integrals for $gg\to HH$ in the limit $m_t^2 \ll s,|t|$ and $m_H=0$.
\item We obtain analytic results for the NLO form factors
  which are used to parametrize the process $gg\to HH$.
  These results can be used to construct the partonic cross section
  in the high-energy limit.
\item We perform an expansion in the Higgs boson mass which converges very
  quickly in the region in which our result is valid. Here the relevant expansion
  parameter is $m_H^2/(2m_t)^2\approx 0.13$. In fact, at LO very good agreement
  with the exact result is obtained after including only the quadratic term.
\item We provide input for the Pad\'e method suggested in Ref.~\cite{Grober:2017uho}
  for the process $gg\to HH$.
\end{itemize}

The remainder of the paper is organized as follows: in Section~\ref{sec::calc}
we describe the method we used to compute the amplitude and master integrals
and discuss the ultraviolet and infra-red structure of the amplitude.
Additionally, we explain our approach to obtain an expansion of the amplitude
in the Higgs boson mass. Afterwards, in Section~\ref{sec::res} we discuss our
results for the form factors and present both analytic and numerical results.
Our conclusions are presented in Section~\ref{sec::con}. In
Appendix~\ref{app::MIs} we define our non-planar master integrals and provide
graphical representations, and in Appendix~\ref{app::BCs} we describe the basis
change which facilitates the computation of the boundary conditions.


\section{\label{sec::calc}Calculation and Renormalization}


\subsection{Non-Planar Master Integrals}

Details on the calculation of the NLO amplitude $gg\to HH$ and in particular
on the reduction to master integrals can be found in
Ref.~\cite{Davies:2018ood}. An algorithm is provided which minimizes
the number of families and yields 10~one-loop and 161~two-loop master
integrals. At one-loop order all integrals are planar. At two-loop order we
obtain 131~planar integrals, which are discussed in detail in~\cite{Davies:2018ood},
and 30~non-planar master integrals. The computation of the latter, which
is based on differential equations, is 
described in the following. A detailed description of the computation of the
boundary conditions can be found in Ref.~\cite{Mishima:2018}.

Graphical representations of the non-planar master integrals can be found in
Appendix~\ref{app::MIs}, see Fig.~\ref{fig::mi2_np}. Note that the 30
non-planar master integrals can be divided into two sets; 16 integrals for
which actual calculation (i.e. solving the differential equations) is
needed, and 14 integrals which can be obtained with the help of crossing
relations. Among the 16 integrals there are 9 seven-line and 7 six-line master
integrals (cf. Fig.~\ref{fig::mi2_np}). We have computed all 30 integrals
directly, however, and use the crossing relations as a cross check.

The main idea to obtain the high-energy expansion is the same as for the
planar integrals; for each integral we make an ansatz which reflects the
expected functional form of the expansion. This ansatz is inserted into the
differential equation obtained by differentiating the master integrals with
respect to $m_t$. It is a new feature of the non-planar integrals that the
ansatz requires both odd and even powers in $m_t$ (see, e.g.,
Ref.~\cite{Kudashkin:2017skd}) whereas for the planar integrals just even
powers were sufficient. Note that due to the structure of the differential
equations w.r.t. $m_t$ the even- and odd-power ansatz terms decouple and can be
treated independently.

For the computation of the planar master integrals in
Ref.~\cite{Davies:2018ood} we followed two approaches. In the first we
computed the boundary integrals in the limit $m_t\to 0$ for a fixed values of
$s$ and $t$ and used differential equations in $t$ to reconstruct the
$t$-dependence (still in the limit $m_t\to 0$). The
differential equations in $m_t$ were then used to construct the expansion
terms in the high-energy limit.  In the second approach $t$-dependent
boundary conditions were computed using asymptotic expansion and Mellin-Barnes
techniques. For the non-planar master integrals we follow only this second
approach, which can be used largely without modification. There are a few
peculiarities, however, mainly connected to the presence of additional regions
in the asymptotic expansion. This requires an extension of the method, which is
described in detail in Ref.~\cite{Mishima:2018}. We note that this method has
many algorithmic elements, which are certainly more generally applicable beyond
the computation of the amplitude described in this paper.

For the computation of the non-planar master integrals (at least for those
with seven lines) it is crucial to choose a basis in which the master
integrals do not contain $\epsilon$ poles in their prefactor in the
amplitude. This guarantees that only the constant ($\epsilon^0$) terms of the
master integrals are required, which contain objects with transcendental
weight of at most four.  We obtain such a basis by replacing dotted
propagators, which are present in the original basis chosen by {\tt
  FIRE}~\cite{Smirnov:2014hma}, with numerator scalar products.  We find a
basis which satisfies the criterion of finite prefactors by testing all
combinations of $G_{j}(1,1,1,1,1,1,1,0,-1)$, $G_{j}(1,1,1,1,1,1,1,-1,0)$,
$G_{j}(1,1,1,1,1,1,1,-1,-1)$, $G_{j}(1,1,1,1,1,1,1,0,-2)$ and
$G_{j}(1,1,1,1,1,1,1,-2,0)$, see Appendix~\ref{app::BCs} for more details.  It
turns out that there is only one such basis within the above candidates.  Our
choice of basis for the $4 \times 4$ and $5 \times 5$ coupled blocks are given
in Appendix~\ref{app::MIs}.

An important cross check of our results is provided by the explicit
expressions from Ref.~\cite{Kudashkin:2017skd} where NLO corrections to Higgs
plus jet were considered in the high-energy limit.  Unfortunately, it is not
possible to simply take over the results from~\cite{Kudashkin:2017skd} since
our amplitude has single poles in $\epsilon$ in the master integral
coefficients if we use their integral basis.  This
means that we would require $\mathcal{O}(\epsilon)$ terms of these master
integrals, which are not known.  We nonetheless compare our results to those
of~\cite{Kudashkin:2017skd}, to the $\epsilon$ orders possible, and they
agree.  Note that the results of~\cite{Kudashkin:2017skd} are given in terms
of kinematics where $t>0, s<0, u<0$, so require analytic continuation to our
physical kinematics.  We have also successfully compared our ``triangle''
master integrals to Ref.~\cite{Anastasiou:2006hc}.  All of our non-planar
results could additionally be cross checked numerically using both {\tt
  FIESTA}~\cite{Smirnov:2015mct} and {\tt pySecDec}~\cite{Borowka:2017idc}.
Analytic results for the master integrals can be found in the ancillary file
to this paper~\cite{progdata_tot}.

In order to illustrate the structure of our results we present the explicit
expression for the pole part of $G_{51}(1,1,1,1,1,1,1,0,-2)$ (see
Appendix~\ref{app::MIs} for the definition of this integral) in the limit
$m_t\to0$.  We include the first and second terms of the small-$m_t$
expansion, and set $s=1$.  The $s$ dependence can easily be restored by making
the replacements $t\to t/s$, $m_t\to m_t/\sqrt{s}$ and multiplying by an
overall factor of $(-\mu^2/s)^{2\epsilon} /s$ to fix the mass dimension of the
integral. Our result reads
\begin{eqnarray}
&&\hspace{-8mm}G_{51}(1,1,1,1,1,1,1,0,-2) =
\nonumber\\&&
\frac{1}{\epsilon} \bigg\{
- \frac{1}{m_t}\frac{2 i \pi^3 \sqrt{-t}}{t \sqrt{1 + t}}
+ \frac{32i\pi - i\pi^3 t(1-t) - 4 t (2 + t) \zeta_3}{2 t (1+t)}
+ \Big(
	\frac{8 (i\pi(1+t) - 2 t)}{t (1+t)}
\nonumber\\&&
	+\frac{8 - i \pi (4+6t+t^2)}{t (1+t)} H_{0}(1+t)
	+\frac{4+2 t+t^2}{2 t (1+t)} {[H_{0}(1+t)]^2}
	-\frac{2 (2+t)^2}{t (1+t)} H_{2}(-t)
\Big) H_{0}(-t)
\nonumber\\&&
+\frac{48(1+t) - \pi^2(6+t) - 24 i \pi t}{3 t (1+t)} H_{0}(1+t)
+ \frac{i \pi (-2+t)}{2 (1+t)} {[H_{0}(1+t)]^2}
\nonumber\\&&
+ \Big(
	\frac{i \pi (2+t)^2}{2 t (1+t)}
	-\frac{(2+t)^2}{2 t (1+t)} H_{0}(1+t)
\Big) {[H_{0}(-t)]^2}
- \frac{2 i \pi (2+3 t)}{t (1+t)} H_{2}(-t)
\nonumber\\&&
- \frac{t}{6 (1+t)} {[H_{0}(-t)]^3}
- \frac{(2+t)^2}{6 t (1+t)} {[H_{0}(1+t)]^3}
- \frac{2 (2+t+t^2)}{t (1+t)} H_{2,1}(-t)
+ \frac{2 (2+t)^2}{t (1+t)} H_{3}(-t)
\nonumber\\&&
+ \left[\frac{t}{1 + t} {[H_{0}(1 + t)]^2}
	+ \left( \frac{16}{1 + t}
	- 2i\pi \frac{(2 + t)^2}{t (1 + t)}
	- \frac{2 (2 + t)^2}{t (1 + t)} H_{0}(1 + t) \right) H_{0}(-t)
\right] \log{(m_t)}
\nonumber\\&&
+ \left[2i\pi \frac{4 + 5 t}{t (1 + t)}
	- \frac{2 t}{1 + t} H_{0}(-t)
	+ \frac{2 (4 + 4 t - t^2)}{t (1 + t)} H_{0}(1 + t)
\right] {\log^2{(m_t)}}
\nonumber\\&&
+ \left[\frac{4 (1 + 2 t)}{3 (1 + t)}\right] {\log^3{(m_t)}}
+ {\cal O}(m_t)
\bigg\}
+ {\cal O}(\epsilon^0)
\,,
\end{eqnarray}
where $H_{\vec{a}}(x)$ denote Harmonic Polylogarithms as defined in~\cite{Remiddi:1999ew}.
Note that here one observes that the leading term is proportional to $1/m_t$;
as explained above, these odd powers of $m_t$ are particular to the non-planar
master integrals and do not appear in the planar results of
Ref.~\cite{Davies:2018ood}.

\begin{figure}[t]
  \centering
    \includegraphics[width=0.95\textwidth]{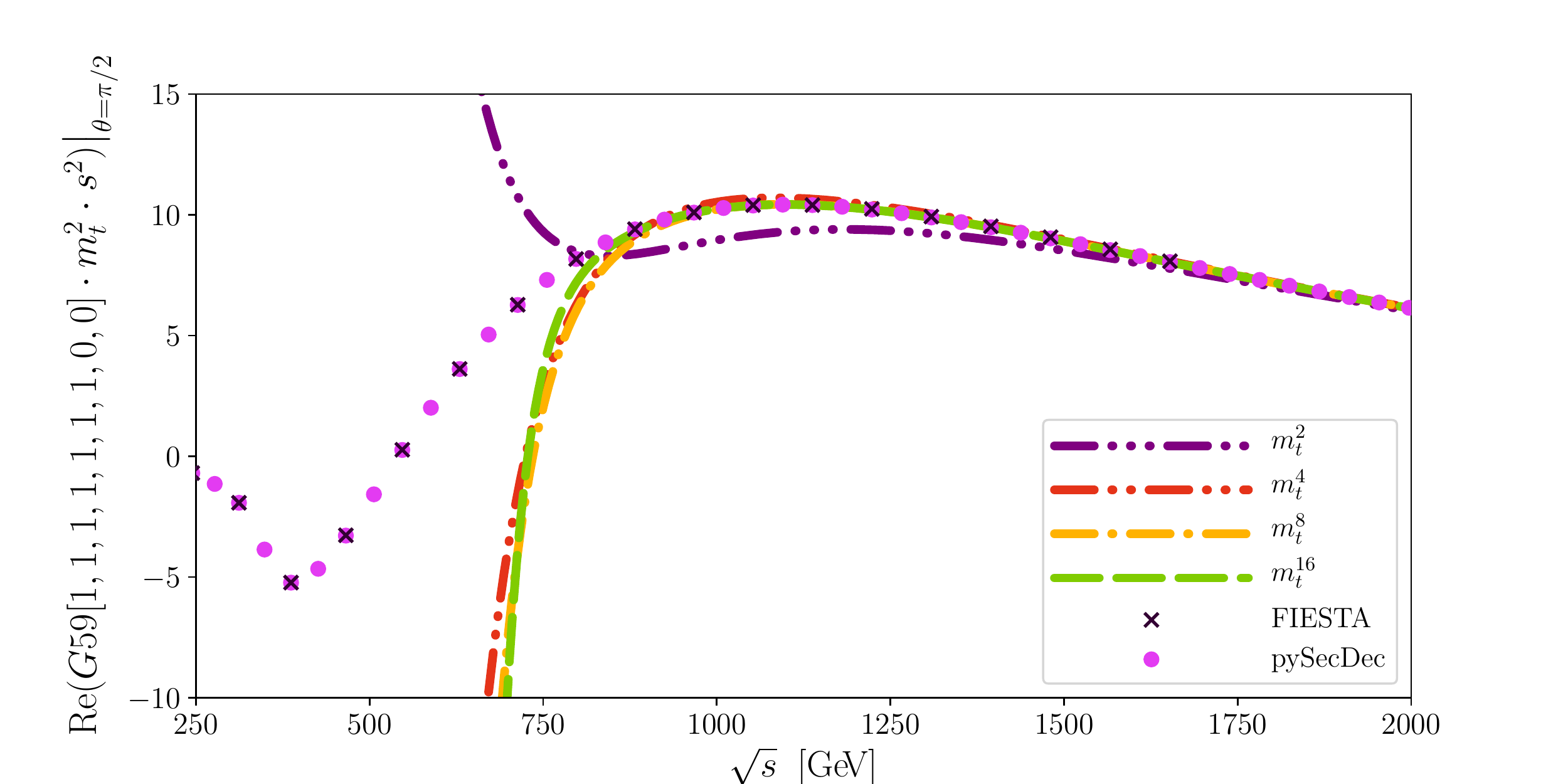}
    \\
    \includegraphics[width=0.95\textwidth]{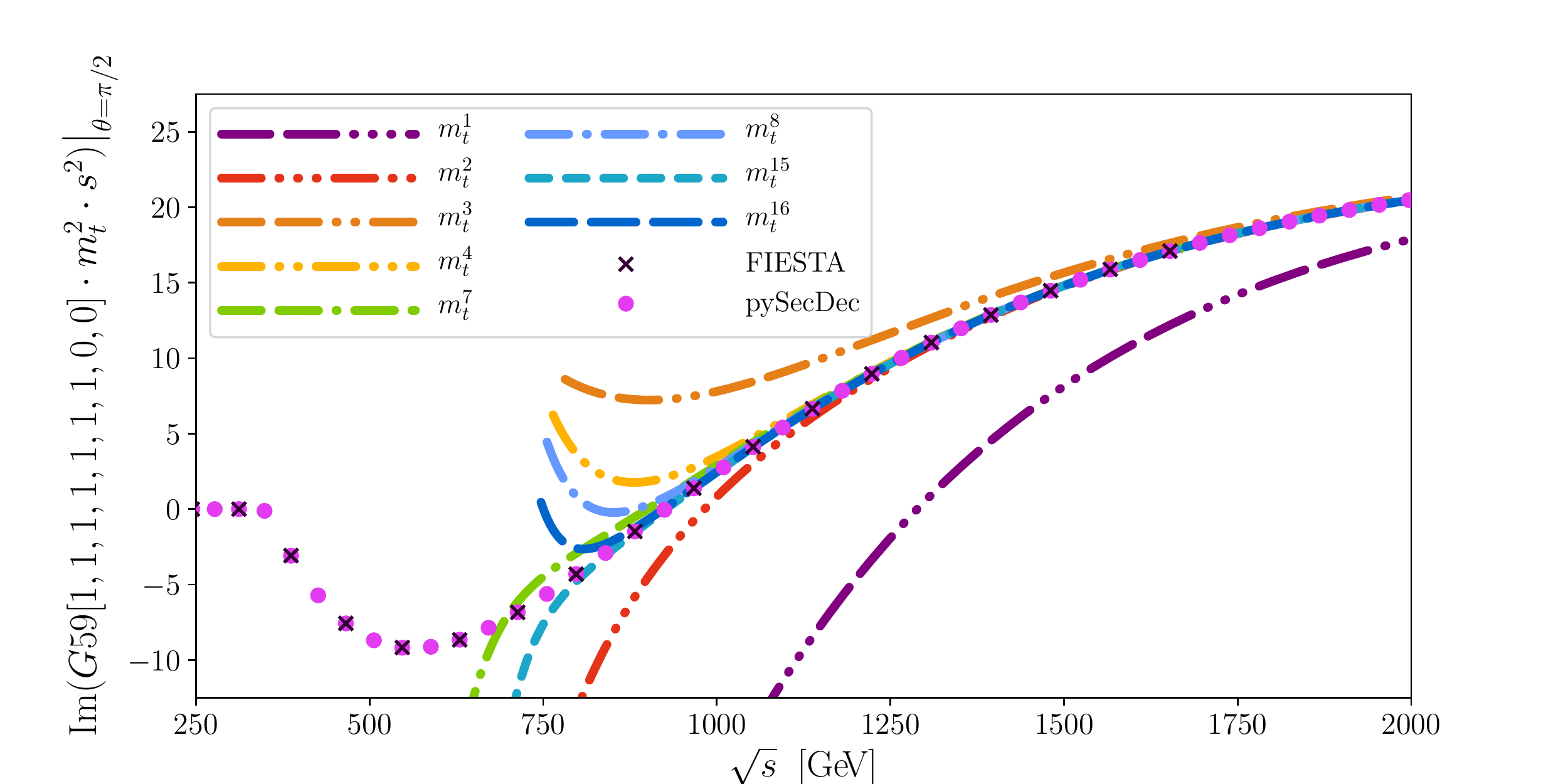}
  \caption{\label{fig::G51}Real and imaginary part of the $\epsilon^0$ term
    of the master integral $G_{59}(1,1,1,1,1,1,1,0,0)$. For clarity we
      rescale by $m_t^2 s^2$.}
\end{figure}

For illustration we show in Fig.~\ref{fig::G51} the real and imaginary part of
the $\epsilon^0$ term of $G_{59}(1,1,1,1,1,1,1,0,0)$ as a function of
$\sqrt{s}$ for $\theta=\pi/2$. We include successively higher orders in the
$m_t$ expansion, which improves the agreement with the exact result shown as
dots ({\tt pySecDec}) and crosses ({\tt FIESTA}).  We want to stress that the
odd $m_t$ terms are numerically significant and are needed to reach the
agreement. It is, furthermore, interesting to mention that after including an
odd expansion term the agreement gets worse and improves only after adding
also the next even $m_t$ term. Thus, if the combination of the $m_t^{2n-1}$
and $m_t^{2n}$ terms are considered a steady improvement is observed. We have
obtained similar plots for all 30 non-planar master integrals.

\subsection{Ultraviolet and Infrared Divergences}

The bare two-loop expressions for the form factors are both ultraviolet and
infrared divergent. We take care of the ultraviolet poles by renormalizing the
top quark mass in the on-shell scheme and the strong coupling constant in the
$\overline{\rm MS}$ scheme using standard one-loop counterterms. Since we
consider the high-energy region we renormalize $\alpha_s$ with six active
quark flavours.

Note that after top quark--mass renormalization the $C_F$ colour factor of the
two-loop form factors are finite. However, there are still infrared
divergences in the $C_A$ colour factor.  We have checked that they agree with
the poles predicted in Ref.~\cite{Catani:1998bh}. We thus construct the
(infrared finite) soft-virtual corrections as
\begin{eqnarray}
  F^{(1)} = F^{(1),\rm IR} - K_g^{(1)} F^{(0)} 
\end{eqnarray}
where $F^{(1),\rm IR}$ is one of the ultraviolet-renormalized, but still
infrared divergent, form factors introduced in Eq.~(\ref{eq::calM}).
$K_g^{(1)}$ can be found in Ref.~\cite{Catani:1998bh}. For the
normalization introduced in Eq.~(\ref{eq::F}) it is given by
\begin{eqnarray}
  K_g^{(1)} &=& - \left(\frac{\mu^2}{-s-i\delta}\right)^\epsilon 
  \frac{e^{\epsilon\gamma_E}}{ 2 \Gamma(1-\epsilon)}
  \left[\frac{C_A}{\epsilon^2} +
    \frac{1}{\epsilon}\left( \frac{11}{6}C_A-\frac{1}{3}n_f \right)
  \right]
  \,,
  \label{eq::Ig1}
\end{eqnarray}
where $\gamma_E$ is Euler's constant.
Note that since infrared and ultraviolet divergences are regulated with the
same parameter $\epsilon$ and since scaleless integrals are set to zero, the
poles in the terms proportional to $n_f$ from Eq.~(\ref{eq::Ig1}) cancel
against the counterterm contribution induced by the $\alpha_s$
renormalization.  However, finite terms proportional to
$\log(\mu^2/(-s-i\delta))$ and the LO result remain. We thus cast $F^{(1)}$ in
the form
\begin{eqnarray}
  F^{(1)} &=& F^{(1),C_F} + F^{(1),C_A} 
              + \beta_0 \log\left(\frac{\mu^2}{-s-i\delta}\right) F^{(0)}
              \,,
\end{eqnarray}
with $\beta_0 = 11 C_A/12 - T n_f/3$. Only $F^{(1),C_F}$ and $F^{(1),C_A}$
contain new information and thus only these will be discussed in the
following. Note that $F^{(1),C_F}$ and $F^{(1),C_A}$ are independent of
$\mu$.


\subsection{\label{sub::expMh}Expansion in $m_H$}

In Ref.~\cite{Davies:2018ood} the calculation has been performed for a
massless Higgs boson which constitutes a good approximation since the relevant
expansion parameter $m_H^2/(2m_t)^2\approx 0.13$ is sufficiently small.  In the
present calculation we incorporate finite Higgs mass effects via an expansion
in $m_H^2/m_t^2$. For our process the dependence on the Higgs boson
mass is analytic, i.e., there are no $\log(m_H)$ terms in the limit $m_H\to0$
since the Higgs boson couples only to the massive top quark. It is thus
possible to perform a simple Taylor expansion (in contrast to a more involved
asymptotic expansion) which we have implemented as follows:
\begin{itemize}
\item We generate the amplitude using the 
  kinematics for a finite Higgs boson mass as given in Eq.~(\ref{eq::q_i^2}).
  In particular, we use $m_H\not=0$ in the projectors onto the individual
  tensor structures and express the amplitude as a linear combination of
  scalar integrals, which depend on $\tilde{s},\tilde{t},m_t$ and $m_H$.
\item Next, the pre-factors of the scalar integrals are expanded about
  $m_H^2 = 0$. Expressions for the Taylor expansion of the scalar integrals
  themselves are constructed using {\tt LiteRed}'s~\cite{Lee:2012cn,Lee:2013mka}
  derivative function {\tt Dinv}.
\item At this point the amplitude is expressed as a linear combination of
  scalar integrals which only depend on $s,t$ and $m_t$; $m_H$ only appears in
  their prefactors. All scalar integrals can be mapped to one of the families
  defined in Ref.~\cite{Davies:2018ood}. We can thus use the same procedure to
  obtain the reduction tables with the help of {\tt FIRE
    5.2}~\cite{Smirnov:2014hma} and {\tt FIRE~5.7}.\footnote{We thank
    Alexander Smirnov for providing us with unpublished versions of {\tt FIRE}
    which we could use to help optimize our reduction.}  Note, however, that
  the number of scalar integrals is significantly increased; at two-loop order
  one has
  about 25,000 scalar integrals to reduce to master integrals, for the $m_H^0$
  contribution. A further 70,000 integrals were reduced in order to produce
  differential equations for the master integrals.  For the $m_H^2$ and
  $m_H^4$ contributions, one must reduce an additional
  123,000 and then
  457,000 integrals respectively.
\end{itemize}
At one-loop order we performed an expansion up to ${\cal O}(m_H^4)$. We show
below that the contribution from the $m_H^4$ terms is very small in the
kinematic region where the small-$m_t$ expansion is valid (see the discussion
regarding Fig.~\ref{fig::XS_1l_mh_norm}).  For this reason, at two loops we
consider only the $m_H^2$ terms of the expansion, and do not perform the
computationally expensive reduction of the above-mentioned additional 457,000
scalar integrals to masters.

The maximum complexities [(number of lines + dots, number of numerators)] of
the integrals appearing in the $m_H^0$ amplitude and differential equations,
in the $m_H^2$ amplitude, and in the $m_H^4$ amplitude are $(10,-4)$, $(9,-5)$
and $(10,-6)$ respectively.



\section{\label{sec::res}Results}


\subsection{Analytic Results for the Form Factors}

In the following we present the leading terms for the three form factors
both in the large-$m_t$ and high-energy limit. We
take the large-$m_t$ term up to order $1/m_t^{12}$
from Ref.~\cite{Grigo:2014jma}.

Using the normalization introduced in Section~\ref{sec::intro} our one-loop
results in the small-$m_t$ limit (showing also the next-to-leading term in the
$m_H$ expansion) is given by
\begin{eqnarray}
  F^{(0)}_{\rm tri} &=& \frac{2 m_t^2}{s}\biggl[4-\lms^2\biggr]
  + {\cal O}\left(\frac{m_t^4}{s^2}\right) \,,\nonumber\\
  F^{(0)}_{\rm box1} &=&  \frac{4m_t^2}{s}
  \biggl[2+\frac{m_H^2}{s}\biggl((\lots-\lts)^2+\pi^2\biggr)\biggr]
  + {\cal O}\left(\frac{m_t^4}{s^2},\frac{m_H^4}{s^2}\right) \,,\nonumber\\
	F^{(0)}_{\rm box2} &=&
		\frac{2 m_t^2}{s t (s+t)}\biggl[
			-\lots^2 (s+t)^2
			-\lts^2 t^2
			-\pi^2 \left(s^2+2 s t+2 t^2\right)
			+ \frac{2 m_H^2}{s(s+t)} \biggl(
				\lots^2 s (s+t)^2
				\nonumber\\&&{}\vph
				+ \pi^2 s^3
				+ 2 s^2 t \left(-2 \lms+\lts+\pi^2-4\right)
				- s t^2 \left(8 \lms+(\lts-2) \lts+16\right)
				\nonumber\\&&{}\vph
				- 4 (\lms+2) t^3
			\biggr)
		\biggr]
	+ {\cal O}\left(\frac{m_t^4}{s^2},\frac{m_H^4}{s^2}\right)\,,
\end{eqnarray}
where
\begin{eqnarray}
  \lms  &=& \log \left(\frac{m_t^2}{s}\right) +i\pi\,,\nonumber\\
  \lts  &=& \log\left(-\frac{t}{s}\right) +i\pi\,,\nonumber\\
  \lots &=& \log\left(1+\frac{t}{s}\right) +i\pi\,.
\end{eqnarray}

For the two-loop form factors we show the coefficients of the $C_F$ and $C_A$
colour factors separately, only to leading order in $m_H$.  In the following,
all symbols $H_{2}$, $H_{3}$, $H_{2,1}$, $H_{4}$, $H_{2,2}$, $H_{2,1,1}$
denote Harmonic Polylogarithms with argument $-t/s$ such that
$H_{2,1,1} = H_{2,1,1}(-t/s)$ etc.
\begin{eqnarray}
  F^{(1),C_F}_{\rm tri} &=& {C_F}\frac{m_t^2}{60 s} 
  \biggl[5 \left(\lms^4-12 \lms^3+144 \lms+240\right)+240 (4 \lms-1) \zeta_3 \nonumber\\
  && +40 \pi^2 \lms (\lms+1)+12 \pi^4\biggr]
  + {\cal O}\left(\frac{m_t^4}{s^2},{\frac{m_t^2m_H^2}{s^2}}\right) \,,\nonumber\\
  F^{(1),C_F}_{\rm box1} &=& {C_F}\frac{m_t^2}{s^3 t (s+t)}
  \biggl[s^2 t^2 \left(12 \lms+\lts (7 \lts+12)+8 \pi^2+20\right)+2 \left(6 \lms+\pi^2+10\right) s^3 t \nonumber\\
  &&{}\vph +\lots^2 (s+t)^2 \left(s^2+6 t^2\right)-12 \lots t (s+t)^2 (\lts t+s)+12 \left(\lts^2+\lts+\pi^2\right) s t^3 \nonumber\\
  &&{}\vph +6 \left(\lts^2+\pi^2\right) t^4+\pi^2 s^4\biggr]
  + {\cal O}\left(\frac{m_t^4}{s^2},{\frac{m_t^2m_H^2}{s^2}}\right)  \,,\nonumber\\
F^{(1),C_F}_{\rm box2} &=& {C_F}\frac{m_t^2}{90 s^3 t (s+t)} 
\biggl[
	30 i \pi  s^2 \Bigl\{6 H_{2} (s+t) \left(s (2 \lots+2 \lts-1)+2 t (\lots+\lts)+t\right)
		\nonumber\\&&{}\vph
		+24 H_{2,1} (s+t)^2
		-12 H_{3} s (s+2 t)
		+2 \lots \left(3 \lts^2+2 \pi^2\right) (s+t)^2
		\nonumber\\&&{}\vph
		+\lts^2 t \left((2 \lts+3) t+6 s\right)
		+\pi^2 \left((1-2 \lts) s^2+2 (3-2 \lts) s t+2 t^2\right)
		\nonumber\\&&{}\vph
		-12 \zeta_3 \left(s^2+2 s t+2 t^2\right)\Bigr\}
	+60 H_{2} s^2 \left(-6 \lots \lts (s+t)^2-3 \lts t (2 s+t)\right.
		\nonumber\\&&{}\vph
		\left.+\pi^2 \left(5 s^2+10 s t+6 t^2\right)\right)
	-180 H_{2,1} s^2 (s+t) \left(s (2 \lots+2 \lts-1)\right.
		\nonumber\\&&{}\vph
		\left.+2 t (\lots+\lts)+t\right)
	-720 H_{2,1,1} s^2 (s+t)^2
	-180 H_{2,2} s^3 (s+2 t)
	\nonumber\\&&{}\vph
	+180 H_{3} s^2 \left(2 \lots (s+t)^2+t (-2 \lts t+2 s+t)\right)
	+720 H_{4} s^2 t^2
	\nonumber\\&&{}\vph
	+90 \lots^2 (s+t)^2 \left(s^2 \left( -3 \lms-\lts^2-\pi^2-7 \right) -3 t^2\right)
	-30 \pi^2 \left(3 s^4 \left(3 \lms-\lts^2+7\right)\right.
		\nonumber\\&&{}\vph
		+s^3 t \left(18 \lms-2 \lts (3 \lts+1)+31\right)
		 +s^2 t^2 (18 \lms-(\lts+5) (3 \lts-8))+18 s t^3
		\nonumber\\&&{}\vph
		\left.{}+9 t^4\right)
	-30 \lts t^2 \left(s^2 (\lts (9 \lms+(\lts-6) \lts+30)+18)+18 (\lts+1) s t+9 \lts t^2\right)
	\nonumber\\&&{}\vph
	-30 \lots^4 s^2 (s+t)^2
	+60 \lots^3 (\lts+3) s^2 (s+t)^2
	+30 \lots \left(\pi^2 s^2 \left((4 \lts+5) s^2 \right.\right.
		\nonumber\\&&{}\vph
		\left.+2 (4 \lts+3) s t+4 (\lts+1) t^2\right)
		+3 t \left((6-2 (\lts-3) \lts) s^3\right.
		\nonumber\\&&{}\vph
		\left.+(12-(\lts-12) \lts) s^2 t+6 (2 \lts+1) s t^2+6 \lts t^3\right)
		\left. +12 s^2 \zeta_3 (s+t)^2\right)
	\nonumber\\&&{}\vph
	-180 s^2 t \zeta_3 (-2 \lts t+2 s+t)
	+\pi^4 s^2 \left(60 s^2+120 s t+73 t^2\right)
	+90 H_{2}^2 s^3 (s+2 t)\biggr]
	\nonumber\\&&{}\vph
	+{\cal O}\left(\frac{m_t^4}{s^2},{\frac{m_t^2m_H^2}{s^2}}\right)\,,
\nonumber\\ 
  F^{(1),C_A}_{\rm tri} &=& {C_A}\frac{m_t^2}{180 s} \biggl[2160-15 \lms^4-60 \left(3+\pi^2\right) \lms^2-2160 (\lms+1) \zeta_3-32 \pi^4\biggr]\nonumber\\
  &&+ {\cal O}\left(\frac{m_t^4}{s^2},{\frac{m_t^2m_H^2}{s^2}}\right)  \,,\nonumber\\
  F^{(1),C_A}_{\rm box1} &=& {C_A}\frac{m_t^2}{s^3 t (s+t)} \biggl[-\lots^2 (s+t)^2 \left(s^2+3 t^2\right)+6 \lots t (s+t)^2 (\lts t+s)\nonumber\\
  &&-\left(4 \lts^2+6 \lts+5 \pi^2-12\right) s^2 t^2-6 \left(\lts^2+\lts+\pi^2\right) s t^3-3 \left(\lts^2+\pi^2\right) t^4-\pi^2 s^4\nonumber\\
  &&-2 \left(\pi^2-6\right) s^3 t\biggr]
  + {\cal O}\left(\frac{m_t^4}{s^2},{\frac{m_t^2m_H^2}{s^2}}\right)  \,,\nonumber\\
  F^{(1),C_A}_{\rm box2} &=& {C_A}\frac{m_t^2}{60 s^3 t (s+t)} \biggl[-10 i \pi  s \Bigl\{6 H_{2} (s+t) \left(s^2 (4 \lots+14 \lts-7)\right.\nonumber\\
  &&{}\vph\left. +s t (4 \lots+14 \lts-17)-4 t^2\right)+48 H_{2,1} s (s+t)^2-84 H_{3} s^2 (s+2 t)\nonumber\\
  &&{}\vph+2 \lots \left(21 \lts^2+19 \pi^2\right) s (s+t)^2+\lts^2 t \left((4 \lts-27) s t-18 s^2+12 t^2\right)\nonumber\\
  &&{}\vph-\pi^2 \left(7 (2 \lts-1) s^3+2 (14 \lts-3) s^2 t+2 (5 \lts+3) s t^2-16 t^3\right)\nonumber\\
  &&{}\vph+12 s \zeta_3 \left(3 s^2+6 s t-4 t^2\right)\Bigr\}-60 H_{2} s \left(-14 \lots \lts s (s+t)^2\right.\nonumber\\
  &&{}\vph\left. +\lts t \left(6 s^2+9 s t-4 t^2\right)+\pi^2 s \left(5 s^2+10 s t+4 t^2\right)\right)\nonumber\\
  &&{}\vph-60 H_{2,1} s (s+t) \left(s^2 (-4 \lots-14 \lts+7)+s t (-4 \lots-14 \lts+17)+4 t^2\right)\nonumber\\
  &&{}\vph+480 H_{2,1,1} s^2 (s+t)^2+420 H_{2,2} s^3 (s+2 t)-60 H_{3} s \left(14 \lots s (s+t)^2\right.\nonumber\\
  &&{}\vph\left. +t \left(-(4 \lts+9) s t-6 s^2+4 t^2\right)\right)-480 H_{4} s^2 t^2+5 \lots^4 s^2 (s+t)^2\nonumber\\
  &&{}\vph-40 \lots^3 \lts s^2 (s+t)^2-10 \lots^2 (s+t) \left(-3 \left((\lts (7 \lts+5)+6) s^3\right.\right.\nonumber\\
  &&{}\vph\left.\left. +(7 \lts (\lts+1)+6) s^2 t+(4 \lts+3) s t^2+3 t^3\right)-19 \pi^2 s^2 (s+t)\right)\nonumber\\
  &&{}\vph-10 \lots \left(\pi^2 s \left((18 \lts-7) s^3+36 \lts s^2 t+9 (2 \lts+1) s t^2-4 t^3\right)\right.\nonumber\\
  &&{}\vph\left. +6 t \left((\lts (4 \lts+3)+3) s^3+(\lts (5 \lts+6)+6) s^2 t+3 (2 \lts+1) s t^2+3 \lts t^3\right)\right.\nonumber\\
  &&{}\vph\left. -36 s^2 \zeta_3 (s+t)^2\right)+5 \lts t^2 \left(\left(\lts^3+54 \lts+36\right) s^2+36 (\lts+1) s t+18 \lts t^2\right)\nonumber\\
  &&{}\vph-60 s t \zeta_3 \left((4 \lts+9) s t+6 s^2-4 t^2\right)-10 \pi^2 \left(3 (\lts (7 \lts-5)-6) s^4\right.\nonumber\\
  &&{}\vph\left. +(42 (\lts-1) \lts-23) s^3 t+(\lts (23 \lts-42)-32) s^2 t^2-2 (4 \lts+9) s t^3-9 t^4\right)\nonumber\\
  &&{}\vph-\pi^4 s^2 \left(195 s^2+390 s t+227 t^2\right) -210 H_{2}^2 s^3 (s+2 t)\biggr]
  + {\cal O}\left(\frac{m_t^4}{s^2},{\frac{m_t^2m_H^2}{s^2}}\right) .
\end{eqnarray}

It is interesting to mention that most of the odd $m_t$ terms, which are
present in the non-planar master integrals, cancel in the amplitude. However,
at higher orders in the $m_t$ expansion there remain odd $m_t$ terms in the
imaginary part of $F^{(1),C_A}_{\rm box1}$ and $F^{(1),C_A}_{\rm box2}$
starting at $m_t^3$.

For completeness we also show the leading terms of the large-$m_t$
expansion which at one-loop order are given by
\begin{eqnarray}
  F^{(0)}_{\rm tri} &=& \frac{4}{3} + {\cal O}(1/m_t^2) \,,\nonumber\\
  F^{(0)}_{\rm box1} &=& -\frac{4}{3} + {\cal O}(1/m_t^2) \,,\nonumber\\
  F^{(0)}_{\rm box2} &=& -\frac{11}{45} \frac{p_T^2}{m_t^2} + {\cal O}(1/m_t^4) \,,
\end{eqnarray}
where
\begin{eqnarray}
  p_T^2 &=& \frac{\tilde{t} \tilde{u} - m_H^4}{\tilde{s}} 
  \,,
\end{eqnarray}
is the (partonic) transverse momentum of the Higgs boson.
At two loops we have
\begin{eqnarray}
  F^{(1),C_F}_{\rm tri} &=& -C_F + {\cal O}(1/m_t^2) \,,\nonumber\\
  F^{(1),C_F}_{\rm box1} &=& C_F + {\cal O}(1/m_t^2) \,,\nonumber\\
  F^{(1),C_F}_{\rm box2} &=& -\frac{131}{810} \frac{p_T^2}{m_t^2} C_F 
  + {\cal O}(1/m_t^4) \,,\nonumber\\ 
  F^{(1),C_A}_{\rm tri} &=& \frac{5}{3} C_A + {\cal O}(1/m_t^2) \,,\nonumber\\
  F^{(1),C_A}_{\rm box1} &=& - \frac{5}{3} C_A + {\cal O}(1/m_t^2) \,,\nonumber\\
  F^{(1),C_A}_{\rm box2} &=& \left[ \frac{308}{675} - \frac{121}{540}\log\left(\frac{-s-i\delta}{m_t^2}\right)\right]\frac{p_T^2}{m_t^2} C_A 
  + {\cal O}(1/m_t^4) \,.
\end{eqnarray}


\subsection{Numerical Results for the Form Factors}

In the following we discuss the $\sqrt{s}$ dependence of the form factors at
one- and two-loop order. If not stated otherwise we use $m_t=173$~GeV and
$m_H=0$ or $m_H=125$~GeV for the top quark and Higgs boson masses,
respectively.

\subsubsection{One-Loop Form Factors}
\begin{figure}[t]
  \centering
  \begin{tabular}{c}
    \includegraphics[width=.95\textwidth]{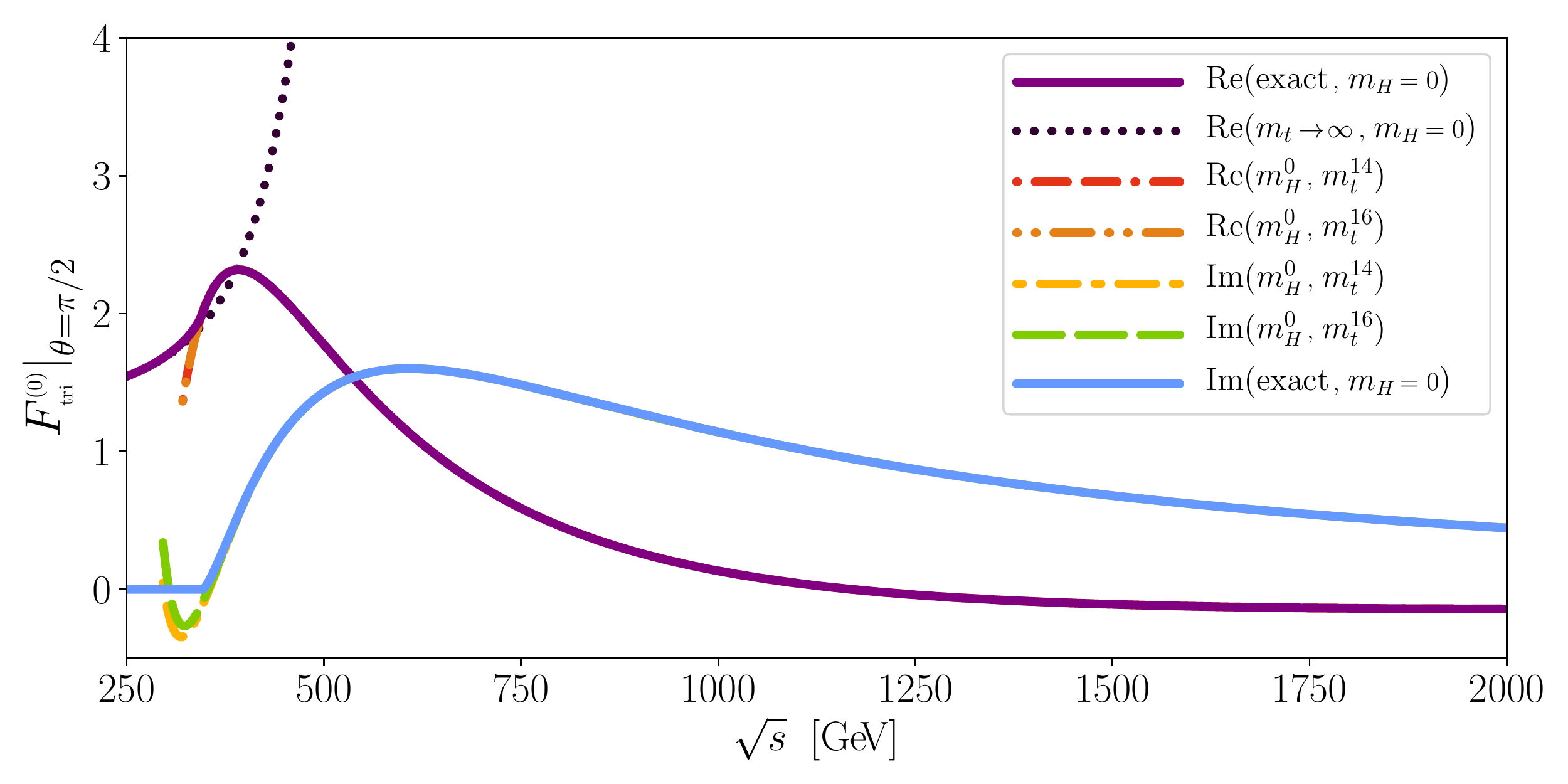}
  \end{tabular}
  \caption{\label{fig::FFtri_1l}The one-loop triangle form factor as a
    function of the partonic center-of-mass energy $\sqrt{s}$ for
    $\theta=\pi/2$. Exact results are shown as solid purple and blue
    curves. The large-$m_t$ expression, which includes terms to $1/m_t^{12}$,
    is the black dotted line. The small-$m_t$ expansions are the dashed lines;
    we show approximations including terms to $m_t^{14}$ and $m_t^{16}$.}
\end{figure}
\begin{figure}[t]
  \centering
  \begin{tabular}{c}
    \includegraphics[width=.95\textwidth]{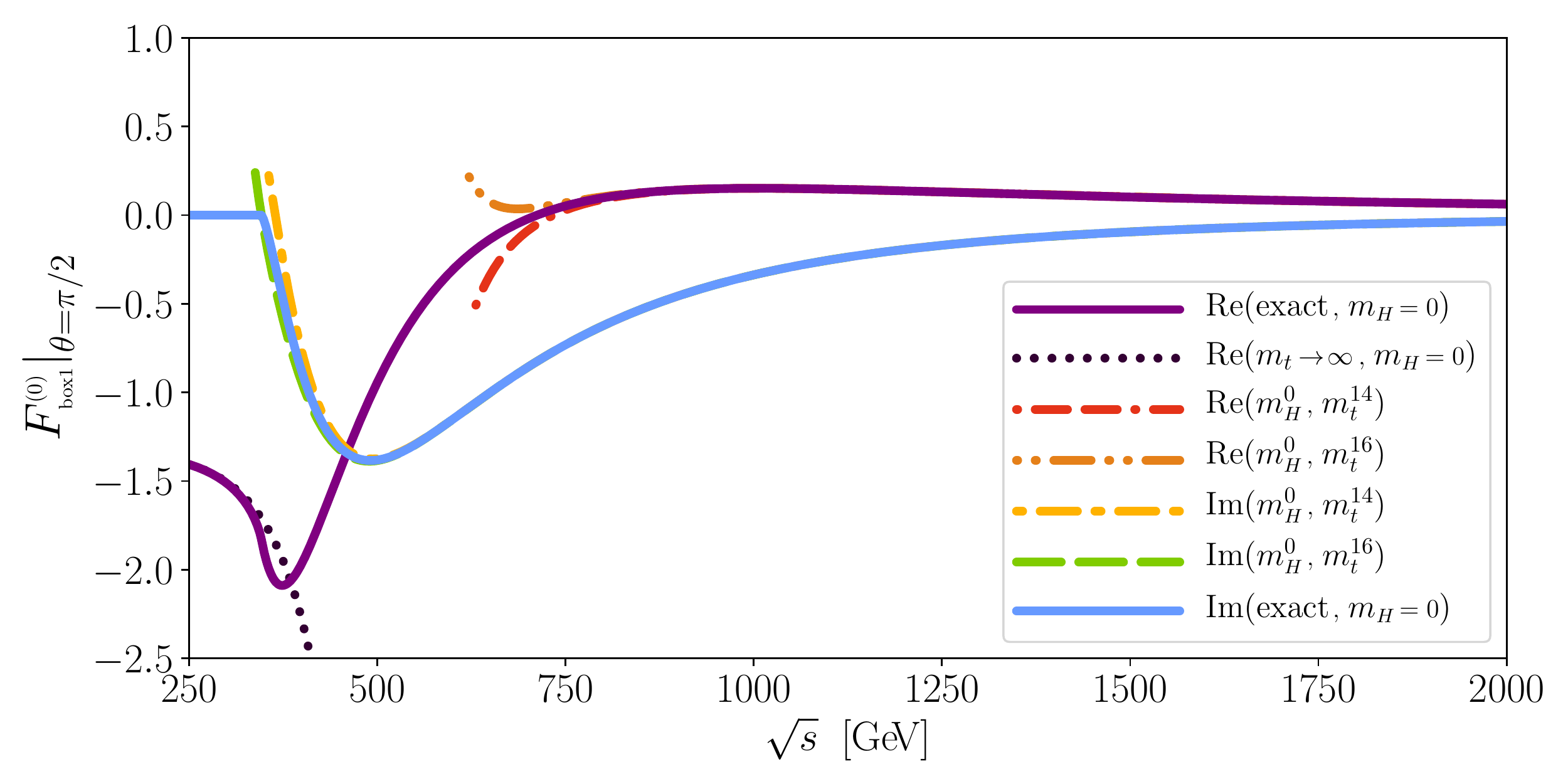} \\
    \includegraphics[width=.95\textwidth]{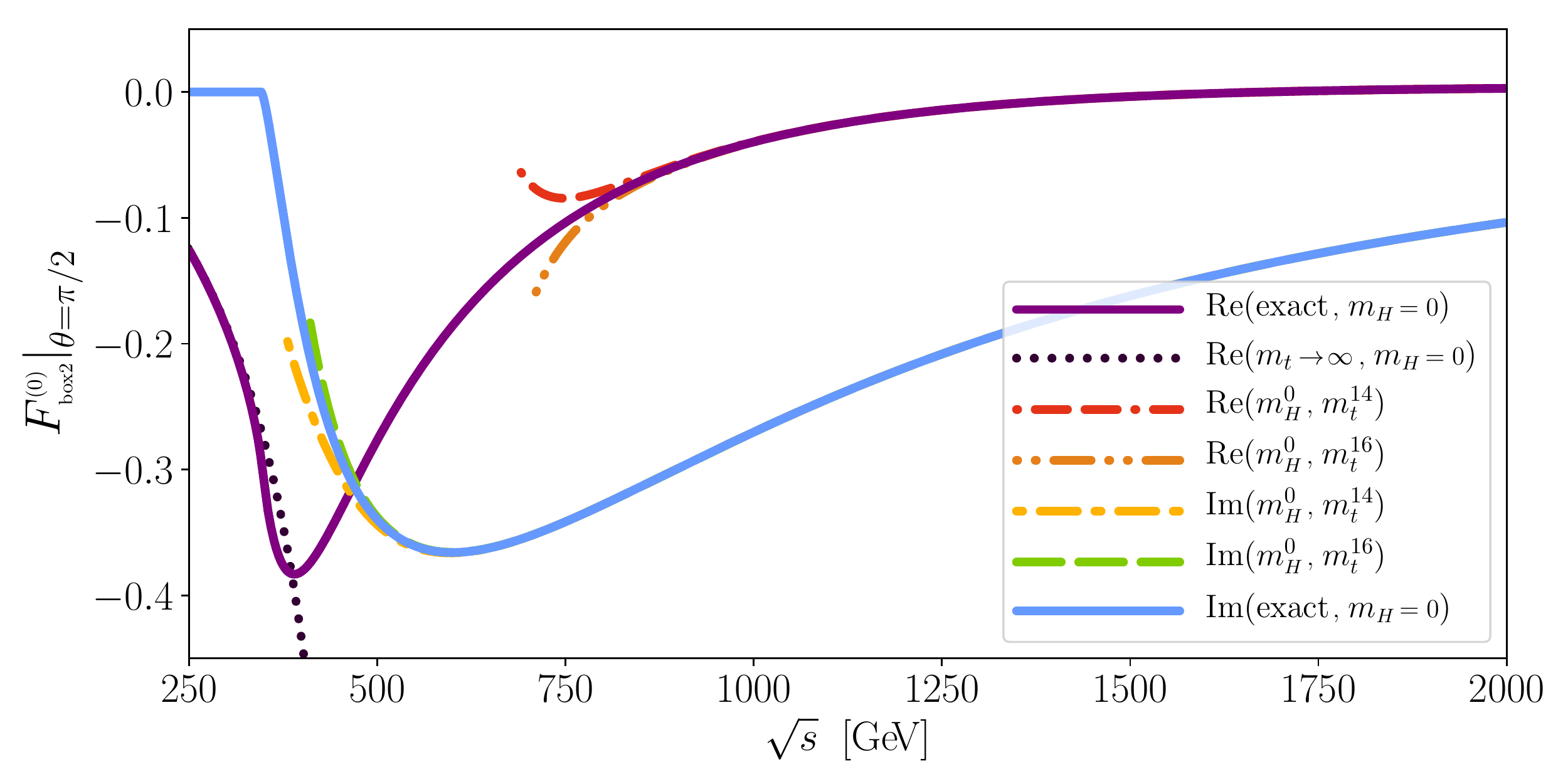} 
  \end{tabular}
  \caption{\label{fig::FFbox_1l}The one-loop box form factors as a function of
    the partonic center-of-mass energy $\sqrt{s}$ for $\theta=\pi/2$. The
    notation is the same as in Fig.~\ref{fig::FFtri_1l}.}
\end{figure}

In Figs.~\ref{fig::FFtri_1l} and \ref{fig::FFbox_1l} we show the one-loop
results where the exact expressions are known and shown as solid curves. Our
high-energy expansions are shown as dashed curves. Both the real and imaginary
parts are plotted.  Note that the imaginary part is zero below
$\sqrt{s}=2m_t$. The large-$m_t$ result is shown as dotted curve.  For the
plots we have chosen $m_H=0$ and $t=-s/2$ which corresponds to a
scattering angle $\theta=\pi/2$ (see Eq.~(\ref{eqn:thetadef})).

The triangle form factor (Fig.~\ref{fig::FFtri_1l}) is approximated very well
by the asymptotic results. The solid and dashed curves lie on top of each
other for the entire $\sqrt{s}$ region above the threshold $2m_t$.  In
Fig.~\ref{fig::FFbox_1l} one observes that for $F^{(0)}_{\rm box1}$ and
$F^{(0)}_{\rm box2}$, the approximations to orders $m_t^{14}$ and $m_t^{16}$
agree with each other, and with the exact result, for values as small as
$\sqrt{s} \approx 800$~GeV and $\sqrt{s} \approx 500$~GeV for the real and
imaginary parts, respectively. Below these energies the curves diverge from
each other.  In general, one can trust the expansions in the regions where
successive approximation orders agree with each other. Due to the very
marginal improvement of the $m_t^{16}$ approximation relative to the
$m_t^{14}$ approximation, we expect that computing higher order terms of the
expansion will not improve the approximation, and that the small-$m_t$
expansion has a finite radius of convergence.

\begin{figure}[t]
  \centering
  \begin{tabular}{c}
    \includegraphics[width=.95\textwidth]{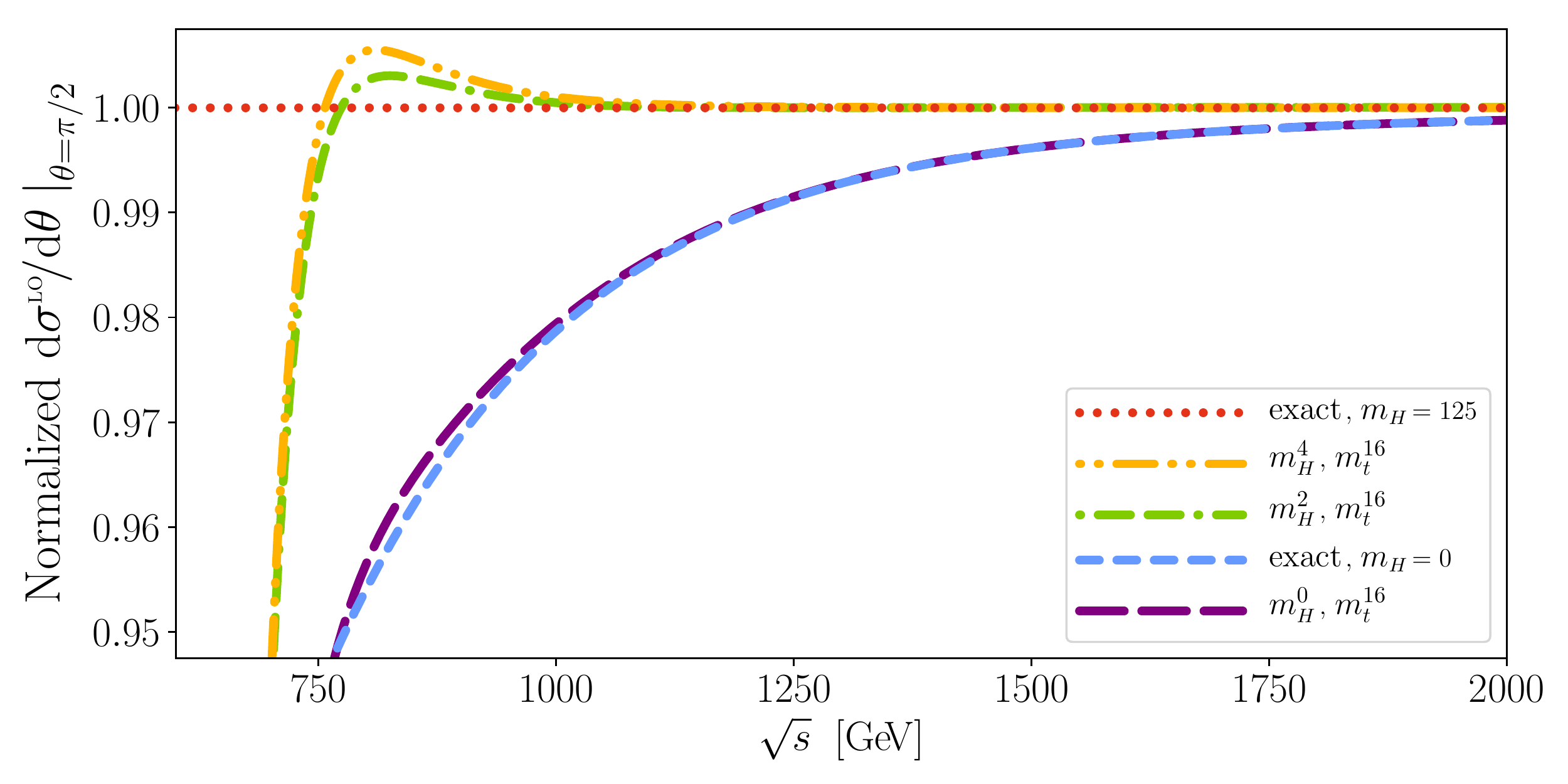}
  \end{tabular}
  \caption{\label{fig::XS_1l_mh_norm} Our approximations to the one loop
    differential cross section. Here we show curves for expansion depths
    $m_H^0$, $m_H^2$ and $m_H^4$. All curves are normalized to the exact
    result, evaluated at $m_H=125$~GeV (red dotted curve).}
\end{figure}

In Fig.~\ref{fig::XS_1l_mh_norm} we consider the $m_H$ dependence of the
partonic cross section for $\theta=\pi/2$. Since this quantity is non-zero for
the whole $\sqrt{s}$ range we can consider the ratio of our approximations to
the exact result, evaluated for $m_H=125$~GeV.  For $\sqrt{s}=1000$~GeV one
observes that the $m_H^0$ approximation (purple dashed curve) reproduces the
$m_H=0$ exact curve well, and that these curves deviate from the $m_H=125$~GeV
exact curve by about 2\%. Including $m_H^2$ terms in the approximation is
sufficient to describe the $m_H=125$~GeV exact curve very well. Including also
$m_H^4$ terms provides a very small correction.  Based on this observation, we
compute $m_H^2$ contributions to NLO quantities but not contributions
proportional to $m_H^4$. We want to remark that the numerical values for
Fig.~\ref{fig::XS_1l_mh_norm} have been obtained by using the
relation
\begin{eqnarray}
  t &\to & m_H^2 - \frac{\tilde{s}}{2}\left(1 - \cos\theta\sqrt{1 -
        \frac{4m_H^2}{\tilde{s}}}\right)
\end{eqnarray}
and performing a consistent expansion in $m_H$. In this way we obtain the 
form factors as a function of $s$, $\theta$ and $m_H$.

\subsubsection{Two-Loop Form Factors}

\begin{figure}
  \centering
  \begin{tabular}{c}
    \includegraphics[width=.95\textwidth]{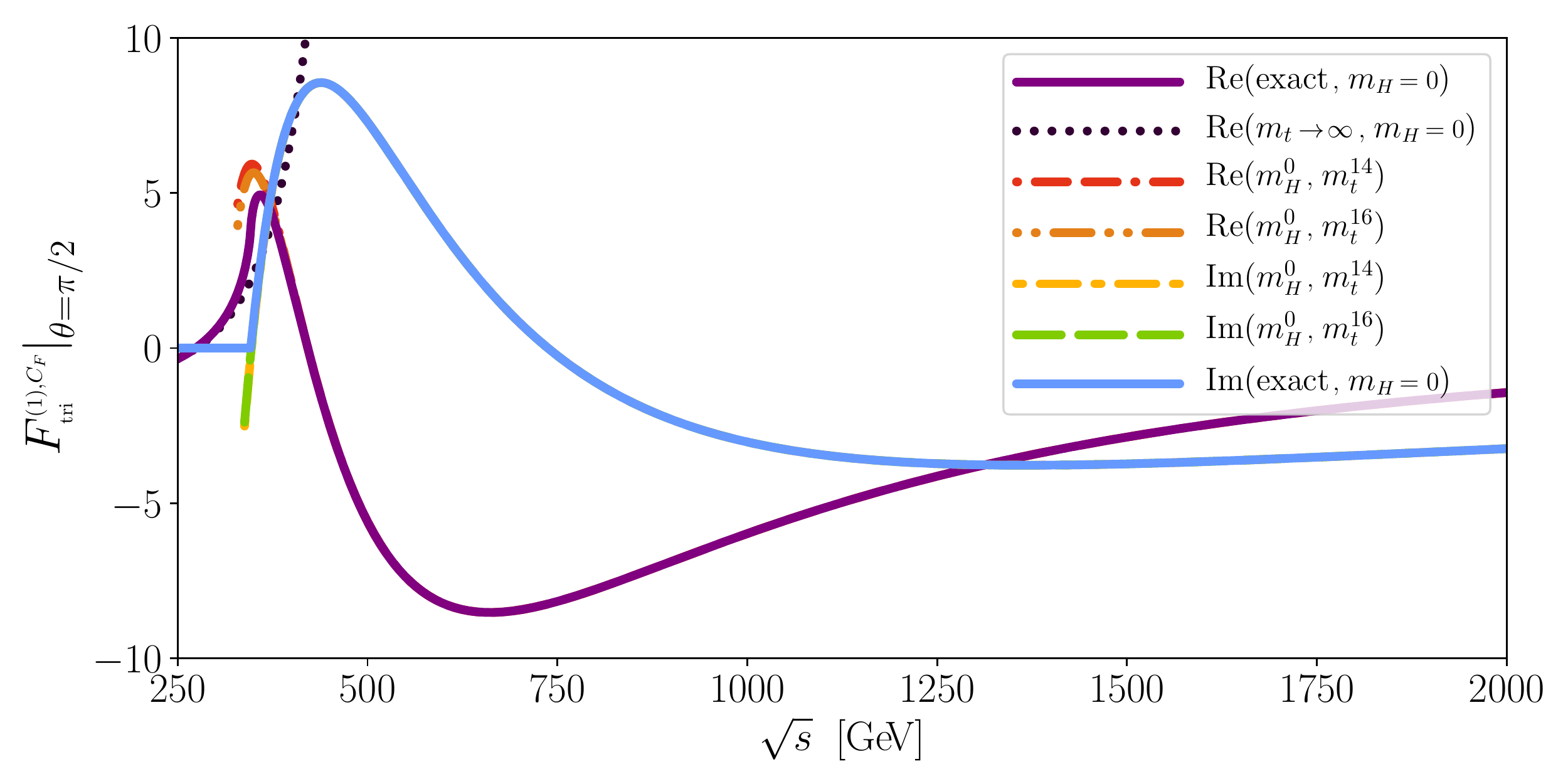}\\
    \includegraphics[width=.95\textwidth]{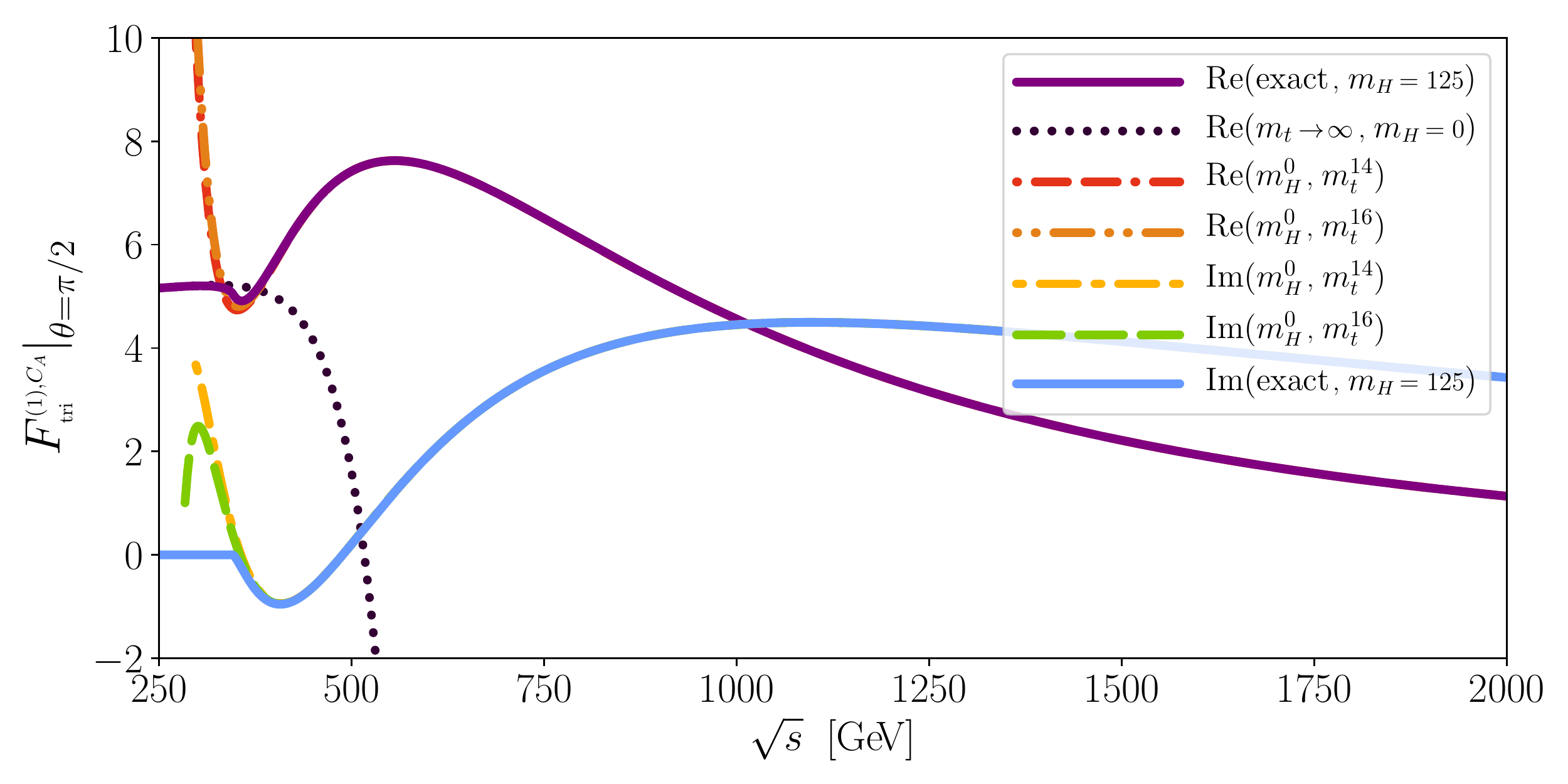}
  \end{tabular}
  \caption{\label{fig::FFtri_2lcf} The two-loop triangle form
    factor $F^{(1),C_F}_{\rm tri}$ as a function of the partonic
    center-of-mass energy for $\theta=\pi/2$.  The same notation as in
    Fig.~\ref{fig::FFtri_1l} is adopted.  We show our approximations (dashed
    curves) for expansion depths $m_t^{14}$ and $m_t^{16}$.}
\end{figure}
\begin{figure}
  \centering
  \begin{tabular}{c}
    \includegraphics[width=.95\textwidth]{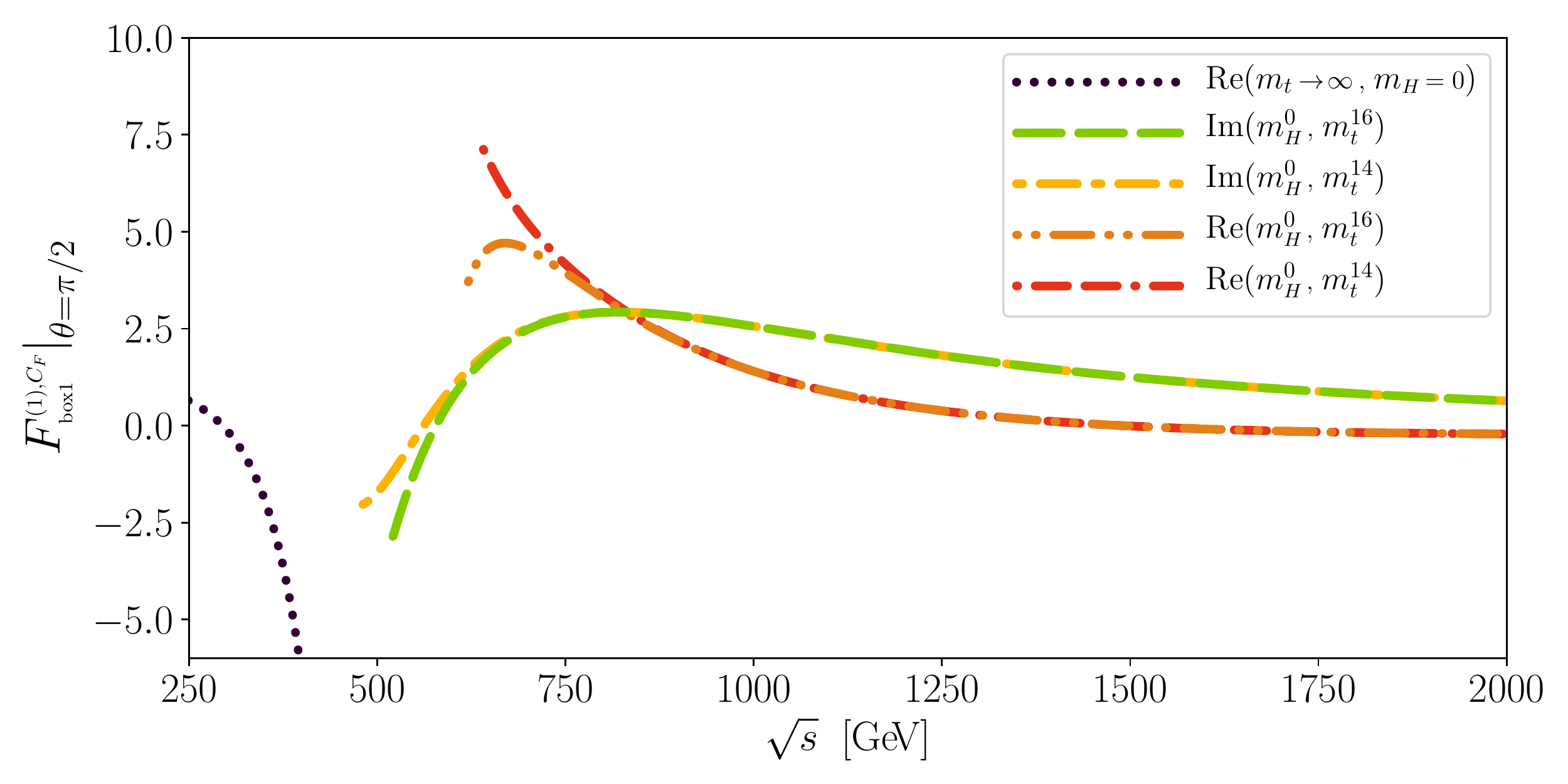} \\
    \includegraphics[width=.95\textwidth]{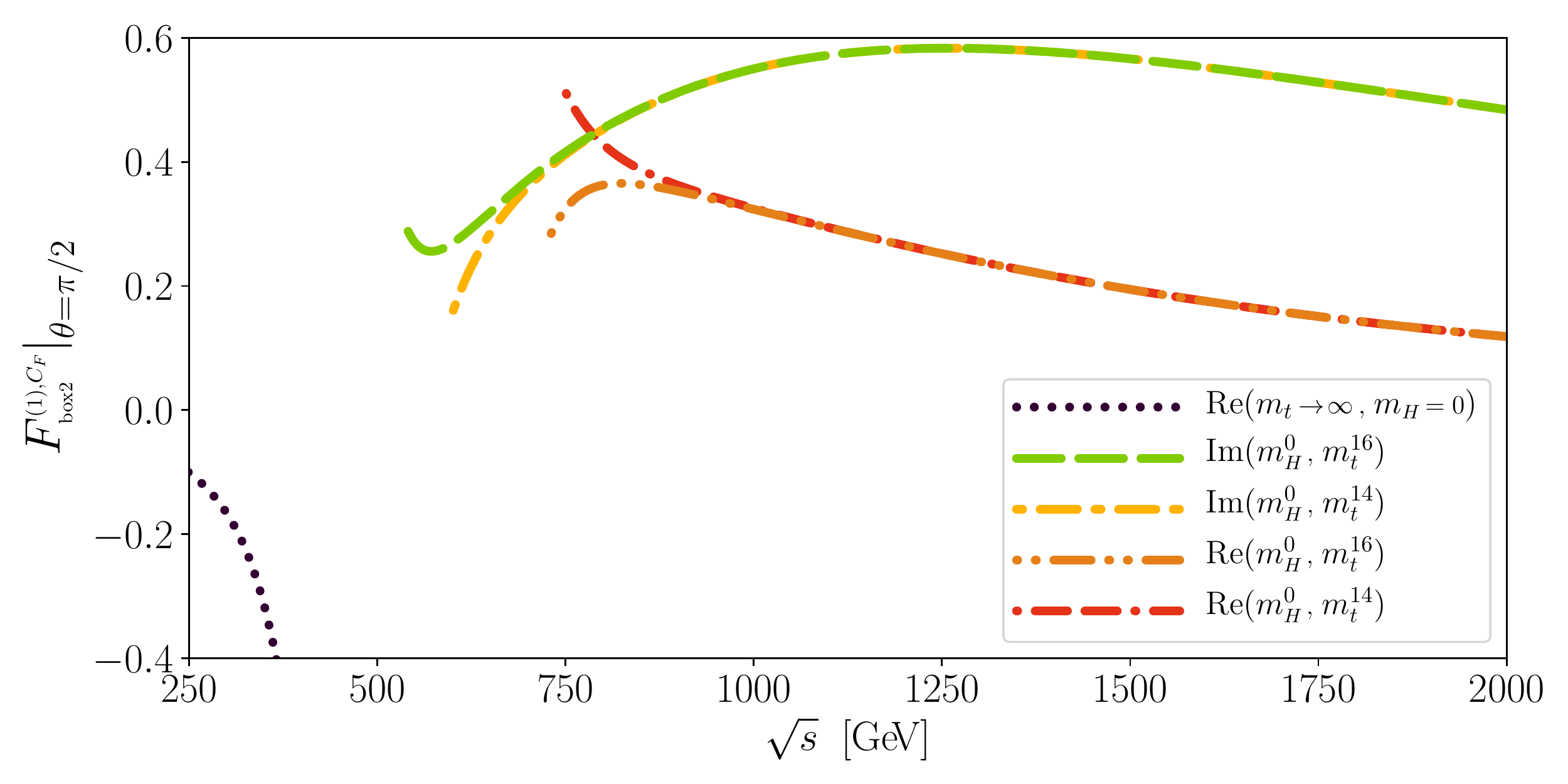} 
  \end{tabular}
  \caption{\label{fig::FFbox_2lcf}The two-loop box form factors
    $F^{(1),C_F}_{\rm box1}$ and $F^{(1),C_F}_{\rm box2}$ as a function of the
    partonic center-of-mass energy for $\theta=\pi/2$.  The same notation as
    in Fig.~\ref{fig::FFtri_1l} is adopted. In these plots the exact curves
    are not known.}
\end{figure}

\begin{figure}
  \centering
  \begin{tabular}{c}
    \includegraphics[width=.95\textwidth]{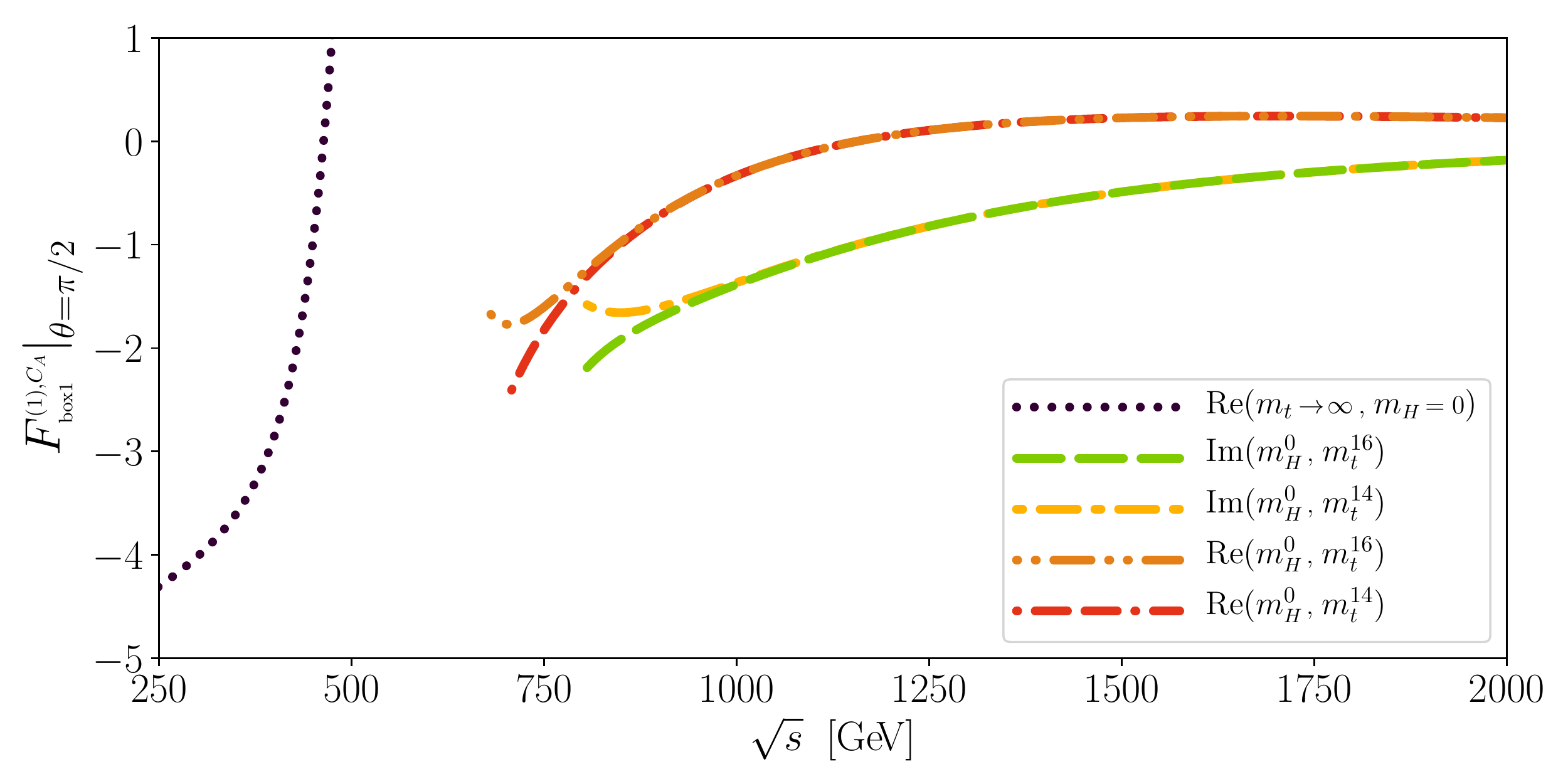} \\
    \includegraphics[width=.95\textwidth]{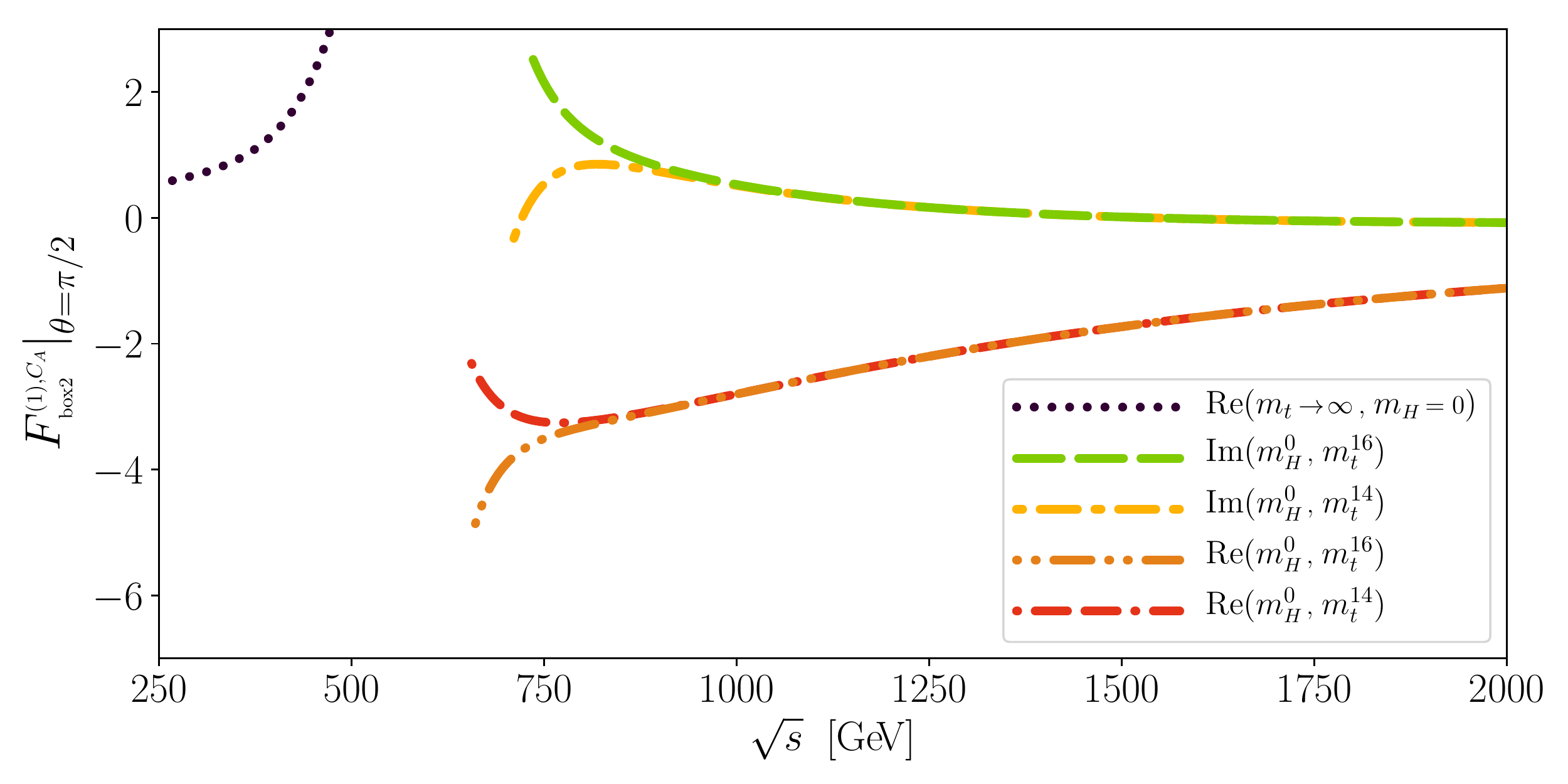} 
  \end{tabular}
  \caption{\label{fig::FFbox_2lca}The two-loop box form factors
    $F^{(1),C_A}_{\rm box1}$ and $F^{(1),C_A}_{\rm box2}$ as a function of the
    partonic center-of-mass energy for $\theta=\pi/2$.  The same notation as
    in Fig.~\ref{fig::FFtri_1l} is adopted. In these plots the exact curves
    are not known.}
\end{figure}

For simplicity we set $m_H=0$ in the following discussion of the two-loop corrections.
The two-loop form factors $F^{(1),C_F}$ and $F^{(1),C_A}$ are shown in
Figs.~\ref{fig::FFtri_2lcf}, \ref{fig::FFbox_2lcf} and~\ref{fig::FFbox_2lca},
where approximations including terms up to $m_t^{14}$ and $m_t^{16}$ are shown.
For the triangle form factor (Fig.~\ref{fig::FFtri_2lcf}) the approximations can
be compared to the exact result from Ref.~\cite{Harlander:2005rq} and, as at
one-loop order, good agreement is found down to $\sqrt{s}\approx 2m_t$.
For the box form factors no exact results are available.
For the $C_F$ contribution we observe a similar behaviour as at one-loop order;
the two highest expansion terms agree down to $\sqrt{s}\approx 800$~GeV and
$\sqrt{s}\approx 500$~GeV for real and imaginary parts respectively, and diverge
for smaller $\sqrt{s}$ values. For the $C_A$ contribution the
convergence properties for real and imaginary part are reversed; we find agreement of
the highest expansion terms down to values $\sqrt{s}\approx 750$~GeV and
$\sqrt{s}\approx 800$~GeV for the real and imaginary parts respectively.


In order to illustrate the size of the $m_H^2$ terms we show in
Tab.~\ref{tab::FF_2l_mh2} for two values of $\sqrt{s}$ the relative
corrections for the real part of the NLO box form factors\footnote{Note that
  the triangle form factors have no non-trivial dependence on $m_H$.} as
compared to the $m_H=0$ result.  One observes corrections up to a few percent,
in agreement with the one-loop results discussed in
Fig.~\ref{fig::XS_1l_mh_norm}.

\newpage

  \begin{table}
    \begin{center}
      \begin{tabular}{c|c|c|c|c}
 \hline
        & $F^{(1),C_F}_{\textnormal{box1}}$ & 
$F^{(1),C_A}_{\textnormal{box1}}$ & $F^{(1),C_F}_{\textnormal{box2}}$ & 
$F^{(1),C_A}_{\textnormal{box2}}$ \\[0.1cm]
 \hline
 $\sqrt{s}=1000$ GeV , $\theta=\pi/2$ & $3.48$ & $-0.30$ & $-5.20$ 
& $1.78$\\
 \hline
 $\sqrt{s}=2000$ GeV , $\theta=\pi/2$ & $1.67$ & $1.26$ & $-0.33$ 
& $0.73$\\
 \hline
 $\sqrt{s}=1000$ GeV , $\theta=\pi/3$ & $4.42$ & $4.26$ & $5.48$ & 
$-0.45$\\
 \hline
 $\sqrt{s}=2000$ GeV , $\theta=\pi/3$ & $2.05$ & $1.11$ & $0.55$ 
& $0.33$\\
 \hline
 \end{tabular}
 \end{center}
 \caption{\label{tab::FF_2l_mh2}
    Correction in percent to the real part of the two-loop form factors 
    induced by $m_H^2$ terms. To obtain the numbers we include the expansion
    in the top quark mass up to $m_t^{16}$.}
\end{table}

\subsubsection{$\theta$ Dependence of the Form Factors}

\begin{figure}
  \centering
  \begin{tabular}{c}
    \includegraphics[angle=90,width=.75\textwidth]{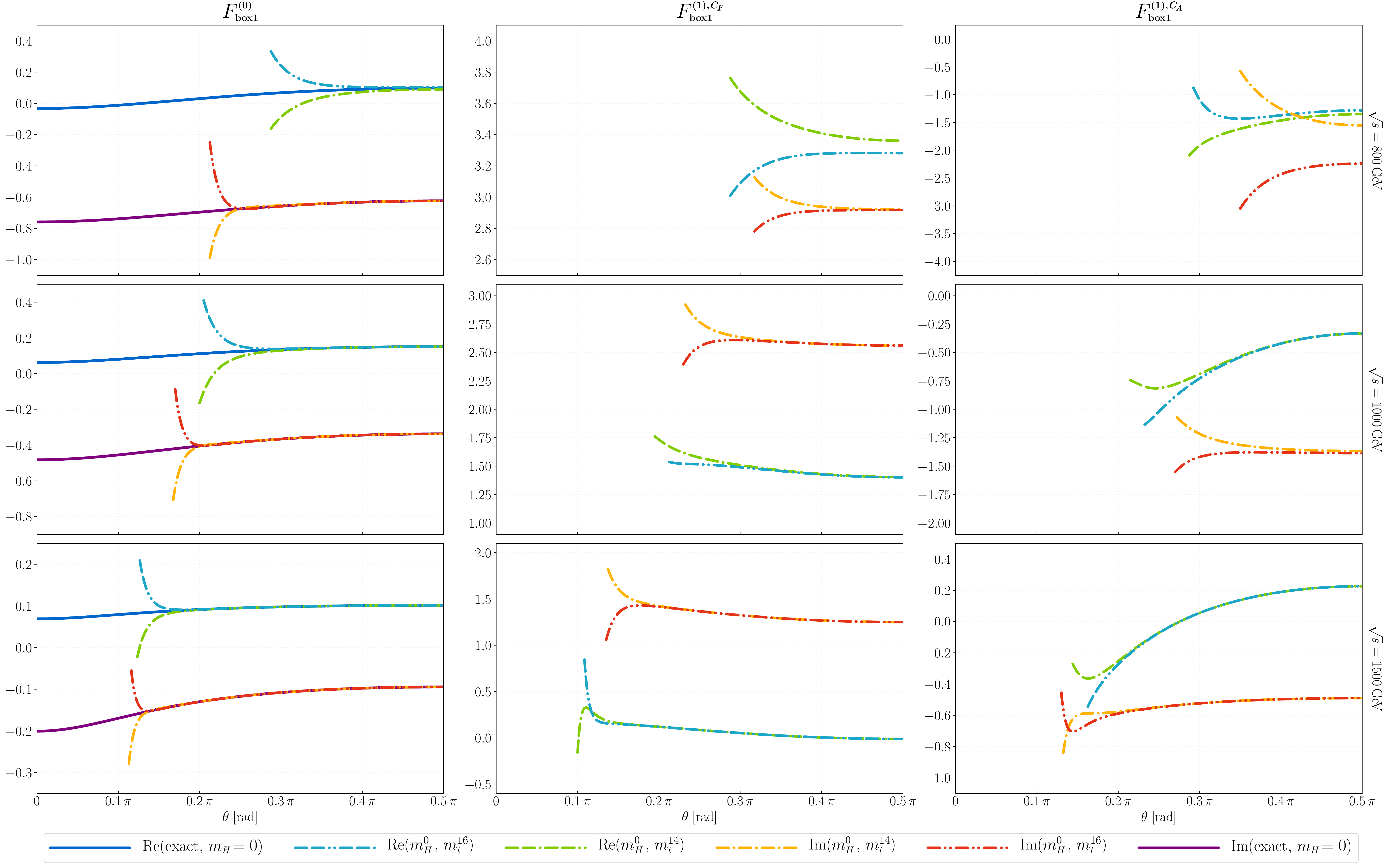} \\
  \end{tabular}
  \caption{\label{fig::FFbox_1_theta}The one- and two-loop box form factors
    $F^{(0)}_{\rm box1}$, $F^{(1),C_F}_{\rm box1}$ and
    $F^{(1),C_A}_{\rm box1}$ (from left to right) as a function of $\theta$
    for three different choices of $\sqrt{s}$ (different columns).
    Both the real and imaginary parts are shown for expansion depths
    $m_t^{14}$ and $m_t^{16}$.}
\end{figure}

\begin{figure}
  \centering
  \begin{tabular}{c}
    \includegraphics[angle=90,width=.75\textwidth]{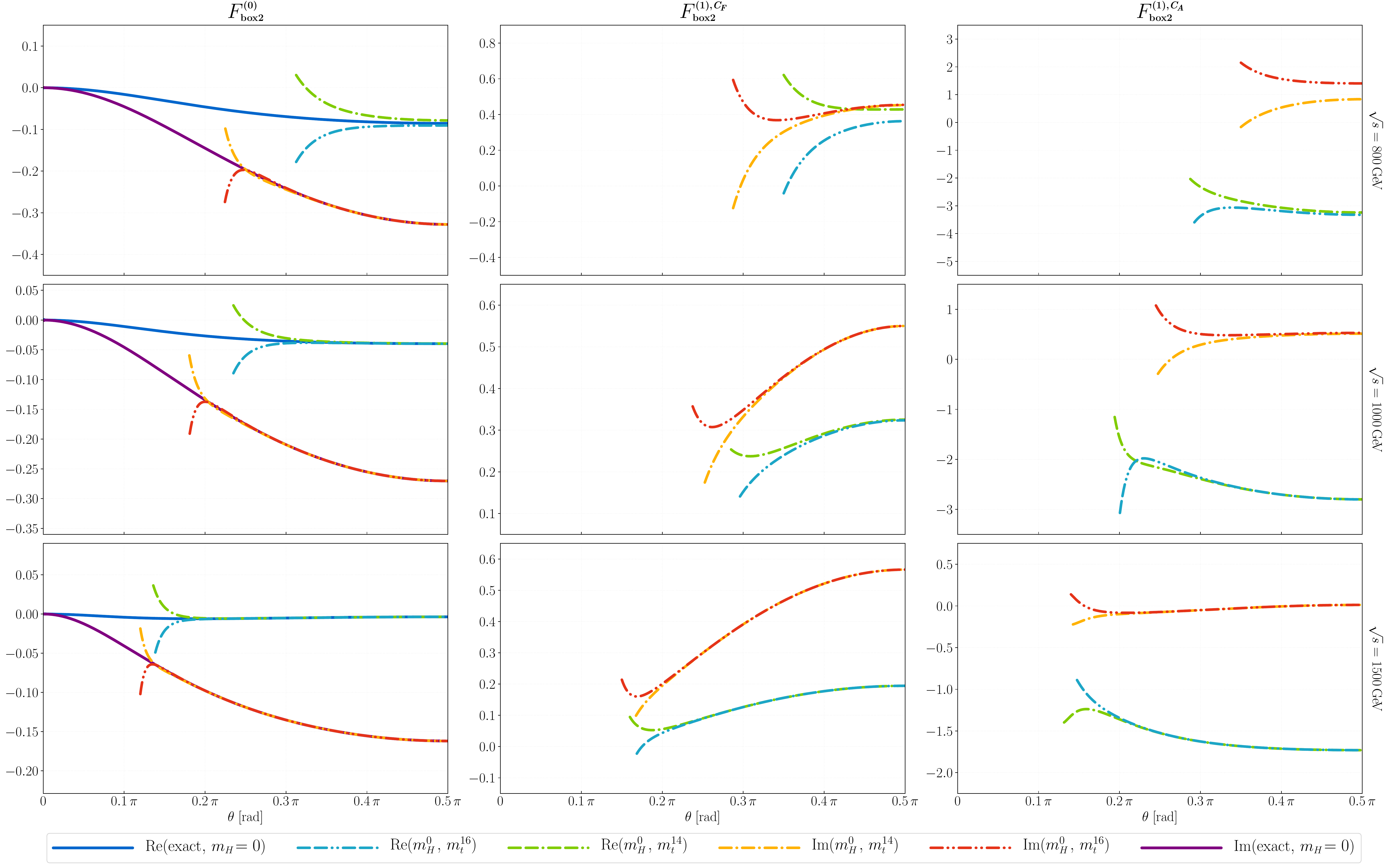} 
  \end{tabular}
  \caption{\label{fig::FFbox_2_theta}As Fig.~\ref{fig::FFbox_1_theta}, but for
    ``box2'' form factors.}
\end{figure}

In the previous subsection we have chosen $\theta=\pi/2$ where $t=-s/2$, i.e.,
the absolute value of $t$ is maximal and our approximation works best.  In
Fig.~\ref{fig::FFbox_1_theta} we show the ``box1'' form factors as a function
$\theta$ with $0\le\theta\le\pi/2$. Symmetric results are obtained for
$\pi/2\le\theta\le\pi$. The form factors $F^{(0)}_{\rm box1}$,
$F^{(1),C_F}_{\rm box1}$ and $F^{(1),C_A}_{\rm box1}$ are shown in the three
columns and the rows correspond to three different choices of $\sqrt{s}$:
$800$~GeV, $1000$~GeV and $1500$~GeV. We show both the real and imaginary part
for expansion depths $m_t^{14}$ and $m_t^{16}$ and assume that our
approximation is good if the two curves agree. At one-loop order we can
compare to the exact result.

In the case of $F^{(0)}_{\rm box1}$ we observe that for
$\sqrt{s}=800$~GeV our approximation works
for values of $\theta$ as low as $0.4\pi$ and $0.25\pi$ for
the real and imaginary part, respectively.
As expected, for larger values of $\sqrt{s}$ the $\theta$ range is
significantly increased; for $\sqrt{s}=1500$~GeV good results
are obtained almost down to $0.1\pi$.

The form factor $F^{(1),C_F}_{\rm box1}$ shows a similar behaviour as
$F^{(0)}_{\rm box1}$. On the other hand, for $F^{(1),C_A}_{\rm box1}$
the $\theta$ range where our approximation works well is 
significantly smaller for $\sqrt{s}=800$~GeV. However, for 
$\sqrt{s}=1000$~GeV and $\sqrt{s}=1500$~GeV similar results
are obtained as for $F^{(0)}_{\rm box1}$ and $F^{(1),C_F}_{\rm box1}$.

Fig.~\ref{fig::FFbox_2_theta} shows analogous results
to Fig.~\ref{fig::FFbox_1_theta} for the ``box2'' for factors.
We observe very similar convergence properties.



\FloatBarrier


\section{\label{sec::con}Conclusions}

We consider Higgs boson pair production in gluon fusion at NLO and compute
analytic results in the high-energy limit where the squared top quark
mass is much smaller than $s$ and $|t|$. We compute analytic results in this
limit for all non-planar master integrals, which complement the results for
the planar integrals, already presented in
Ref.~\cite{Davies:2018ood}. Analytic expressions for the master integrals are
provided in an ancillary file to this paper \cite{progdata_tot}.  The results
are used to obtain analytic expressions for the form factors of the $gg\to HH$
amplitude, including expansion terms up to $m_t^{16}$.  For large scattering
angles (which means large $|t|$) we show that our calculation provides good
approximations for $\sqrt{s}$ values down to about $700$ to $800$~GeV.  Finite
Higgs boson mass corrections are incorporated as an expansion in
$m_H^2/m_t^2$, which converge quickly in the regions where we have
$m_t^2\ll s,|t|$.

Our expressions allow for a fast numerical evaluation of the form factors and
thus provide an alternative to the exact, numerically expensive calculation of
Ref.~\cite{Borowka:2016ypz} in the high-energy region of the
phase-space. It is in particular tempting to combine our results with other
approximations~\cite{Grigo:2014jma,Degrassi:2016vss,Grober:2017uho,Bonciani:2018omm}
to cover the full phase space. Such investigations are the subject of ongoing
research.

 
\section*{Acknowledgements}

We thank Alexander Smirnov and Vladimir Smirnov for many useful
discussions. We are grateful to Christopher Wever for fruitful discussions and
for providing results from Ref.~\cite{Kudashkin:2017skd} which are not
publicly available.  We also thank Hjalte Frellesvig for discussions on uniform
transcendental bases.  D.W. acknowledges the support by the DFG-funded
Doctoral School KSETA.  This work is supported by the BMBF grant 05H15VKCCA.



\begin{appendix}


\section{\label{app::MIs}Non-Planar Master Integrals at Two Loops}

Altogether we encounter 10 one-loop and 161 two-loop master integrals; 30 of
the latter are non-planar.  The definitions of all one- and two-loop integrals
and the graphical representations of the one- and planar two-loop master
integrals can be found in Ref.~\cite{Davies:2018ood}.  In the following we
provide the complementary information for the 30 non-planar integrals.

\begin{figure}[b]
  \centering
  \includegraphics[width=\textwidth]{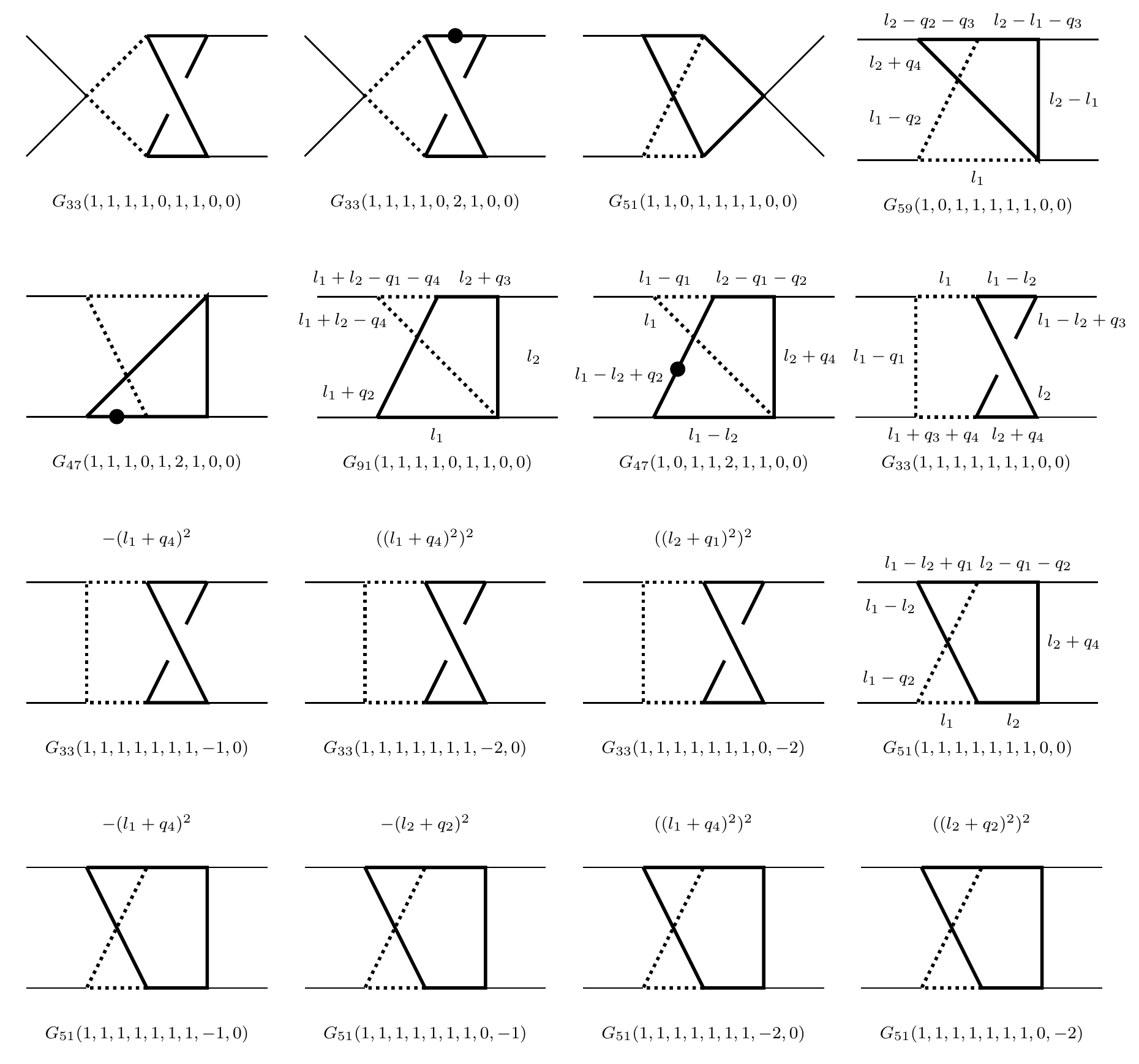}
  \caption{\label{fig::mi2_np}Sixteen two-loop non-planar master
    integrals. Solid and dashed lines represent massive and massless scalar
    propagators, respectively. The external (thin) lines are massless.
    Squared propagators are marked by a dot and numerators are explicitly
    given above the diagrams (see also the definitions of the families in
    Eq.~(\ref{eq::D_ij})). The remaining 14 non-planar master integrals, which
    are not shown, are obtained by crossing.}
\end{figure}

It is easy to see that two-loop integrals with five lines or fewer are
all planar and thus the two-loop non-planar integrals have either six or
seven lines.  Due to crossing symmetries it is sufficient to solve the
differential equations only for
the 16 integrals shown in Fig.~\ref{fig::mi2_np}; the analytic results
for the remaining 14 integrals can be obtained by applying the crossing relations
$s\leftrightarrow t$, $s\leftrightarrow u$ or $t\leftrightarrow u$.

Altogether we use five integral families to accommodate all 30
integrals. They are defined in the following way,
\begin{align}
D_{33}(q_1,q_2,q_3,q_4)&=\left\{
-l_1^2,
m_t^{2}-l_2^2,
m_t^{2}-(l_2+q_4)^2,
-(l_1+q_3+q_4)^2,
-(l_1-q_1)^2,
\right.\nonumber\\
&\left.
m_t^{2}-(l_1-l_2+q_3)^2,
m_t^{2}-(l_1-l_2)^2,
-(l_1+q_4)^2,
-(l_2+q_1)^2
\right\}\,,\nonumber\\
D_{47}(q_1,q_2,q_3,q_4)&=\left\{
-l_1^2,
m_t^{2}-l_2^2,
m_t^{2}-(l_2+q_4)^2,
m_t^{2}-(l_2-q_1-q_2)^2,
\right.\nonumber\\&\left.
m_t^{2}-(l_1-l_2+q_2)^2,
m_t^{2}-(l_1-l_2)^2,
-(l_1-q_1)^2,
-(l_1+q_4)^2,
\right.\nonumber\\&\left.
-(l_2+q_1)^2
\right\}\,,\nonumber\\
D_{51}(q_1,q_2,q_3,q_4)&=D_{47}(q_2,q_1,q_3,q_4)\,,\nonumber\\
D_{59}(q_1,q_2,q_3,q_4)&=D_{47}(q_2,q_3,q_1,q_4)\,,\nonumber\\
D_{91}(q_1,q_2,q_3,q_4)&=\left\{
m_t^{2}-l_1^2,
m_t^{2}-(l_1+q_2)^2,
-(l_1+l_2-q_4-q_1)^2,
-(l_1+l_2-q_4)^2,
\right.\nonumber\\
&\left.
m_t^{2}-(l_1-q_4)^2,
m_t^{2}-(l_2+q_3)^2,
m_t^{2}-l_2^2,
-(l_2+q_2)^2,
-(l_2+q_4)^2
\right\}\,,
\label{eq::D_ij}
\end{align}
where $l_1$ and $l_2$ are the loop momenta.
The complete set of two-loop non-planar master integrals is then given by
{\scalefont{0.70}
  \begin{align}
    \begin{array}{lllll}
      G_{33}(1, 1, 1, 1, 0, 1, 1, 0, 0),
      &G_{33}(1, 1, 1, 1, 0, 2, 1, 0, 0),
      &G_{33}(1, 1, 1, 1, 1, 1, 1, 0, 0),
      &G_{33}(1, 1, 1, 1, 1, 1, 1, -1, 0),\\
      G_{33}(1, 1, 1, 1, 1, 1, 1, -2, 0),
      &G_{33}(1, 1, 1, 1, 1, 1, 1, 0, -2),
      &G_{47}(1, 0, 1, 1, 2, 1, 1, 0, 0),
      &G_{47}(1, 1, 1, 0, 1, 2, 1, 0, 0),\\
      G_{51}(1, 1, 0, 1, 1, 1, 1, 0, 0),
      &G_{51}(1, 1, 1, 1, 1, 1, 1, 0, 0),
      &G_{51}(1, 1, 1, 1, 1, 1, 1, -1, 0),
      &G_{51}(1, 1, 1, 1, 1, 1, 1, 0, -1),\\
      G_{51}(1, 1, 1, 1, 1, 1, 1, -2, 0),
      &G_{51}(1, 1, 1, 1, 1, 1, 1, 0, -2),
      &G_{59}(1, 0, 1, 1, 1, 1, 1, 0, 0),
      &G_{59}(1, 1, 0, 1, 1, 1, 1, 0, 0),\\
      G_{59}(1, 1, 1, 1, 1, 1, 1, 0, 0),
      &G_{59}(1, 1, 1, 1, 1, 1, 1, -1, 0),
      &G_{59}(1, 1, 1, 1, 1, 1, 1, 0, -1),
      &G_{59}(1, 1, 1, 1, 1, 1, 1, -2, 0),\\
      G_{59}(1, 1, 1, 1, 1, 1, 1, 0, -2),
      &G_{91}(0, 1, 1, 1, 1, 1, 1, 0, 0),
      &G_{91}(1, 0, 1, 1, 1, 1, 1, 0, 0),
      &G_{91}(1, 0, 1, 1, 1, 1, 2, 0, 0),\\
      G_{91}(1, 1, 1, 1, 0, 1, 1, 0, 0),
      &G_{91}(1, 1, 1, 1, 1, 1, 1, 0, 0),
      &G_{91}(1, 1, 1, 1, 1, 1, 1, -1, 0),
      &G_{91}(1, 1, 1, 1, 1, 1, 1, 0, -1),\\
      G_{91}(1, 1, 1, 1, 1, 1, 1, -2, 0),
      &G_{91}(1, 1, 1, 1, 1, 1, 1, 0, -2)\,.
      &&&
    \end{array}
          \nonumber
  \end{align}
  }
\vspace{-8mm}
\begin{equation}
\label{eq::MI2l}
{}
\end{equation}
Note that at two-loop order, each family is defined using seven propagators
and two irreducible numerators which correspond to the last two indices.

We present analytic results for all integrals in Eq.~(\ref{eq::MI2l}) as an
expansion for $m_t^2\ll s,|t|$ in the ancillary file to this
paper~\cite{progdata_tot}. For the integration measure we use
$(\mu^2)^{(4-d)/2}e^{\epsilon\gamma_E}{\rm d}^d k/{ (i\pi^{d/2}) }$ where
$d=4-2\epsilon$ is the space-time dimension.


\section{\label{app::BCs}Non-Planar Master Integral Basis}

For the six-line non-planar master integrals all boundary conditions can be
computed for the original {\tt FIRE} basis.  We use the method described in
detail in~\cite{Mishima:2018}.

For the seven-line non-planar integrals we first rewrite the integrals with
dots (in the following denoted by a superscript ``(d)'') in terms of the
integrals with numerators (superscript ``(n)'') using integration-by-parts
relations.  The latter are chosen such that the amplitude has no $\epsilon$
poles in the prefactors of the integrals.

Altogether we have 19 seven-line master integrals which decompose into
a $4\times4$ and three $5\times5$ blocks.
For illustration we briefly discuss the $4\times4$ block of family $G_{33}$,
where the relation between the integrals reads
\begin{align}
{\scalefont{0.6}
  \vec{I}_{33}^{\,(n)}=
  \left(
  \begin{array}{cccc}
    1&0&0&0\\
    \!\!\!\frac{st}{s+2t}
    +
    m_t^2\left( \frac{-4s}{s+2t}+\epsilon\frac{8s}{s+2t} \right)
    + {\cal O}(m_t^4,\epsilon^2)
     &
       m_t^0(\ldots)
       &m_t^2(\ldots)&m_t^2(\ldots)\\
    m_t^0(\ldots) &m_t^0(\ldots) &m_t^2(\ldots)&m_t^2(\ldots)\\
    m_t^0(\ldots) &m_t^0(\ldots) &m_t^2(\ldots) &
                          m_t^2\left( -\frac{st}{4\epsilon}
                          -\frac{1}{2}(s+t)(3s+2t)
                          + {\cal O}(m_t^4,\epsilon)
                          \right)
  \end{array}
                          \right)
                          \vec{I}_{33}^{\,(d)}
}                          \nonumber\\
                          +\mbox{simpler integrals}
                          \,,
                          \label{eq:basis-change33}
\end{align}
with
\begin{align}
\vec{I}_{33}^{\,(n)}=
\left(
\begin{array}{l}
G_{33}(1, 1, 1, 1, 1, 1, 1, 0, 0)\\
G_{33}(1, 1, 1, 1, 1, 1, 1, -1, 0)\\
G_{33}(1, 1, 1, 1, 1, 1, 1, -2, 0)\\
G_{33}(1, 1, 1, 1, 1, 1, 1, 0, -2)
\end{array}
\right)
,\quad
\vec{I}_{33}^{\,(d)}=
\left(
\begin{array}{l}
G_{33}(1, 1, 1, 1, 1, 1, 1, 0, 0)\\
G_{33}(1, 1, 1, 1, 2, 1, 1, 0, 0)\\
G_{33}(1, 1, 2, 1, 1, 1, 1, 0, 0)\\
G_{33}(1, 2, 1, 1, 1, 1, 1, 0, 0)
\end{array}
\right)
\,.
\label{eq:InId33}
\end{align}
In Eq.~(\ref{eq:basis-change33}) we only show some of the matrix elements; the
others have a similar structure.

To obtain the finite $(m_t^2/s)^0$ terms for the four integrals of
$\vec{I}_{33}^{\,(n)}$ we must compute the coefficients of the leading terms
in the small-$m_t$ limit of $\vec{I}_{33}^{\,(d)}$.  In practice, that is the
coefficients of $(m_t^2/s)^{-1/2}$ and $(m_t^2/s)^0$ for the first entry, and
for the second entry the coefficients of $(m_t^2/s)^{-1}$, $(m_t^2/s)^{-1/2}$
and $(m_t^2/s)^{0}$. For the third and fourth entries, the coefficients of
$(m_t^2/s)^{-3/2}$ and $(m_t^2/s)^{-1}$ are needed.  All other higher order
terms need not be computed for the boundary conditions.

By inspecting the matrix in Eq.~(\ref{eq:basis-change33}) one observes that
for the ${\cal O}(m_t^0)$ terms at most the constant term in the $\epsilon$
expansion has to be computed. All $1/\epsilon$ poles
in~(\ref{eq:basis-change33}) are suppressed by a factor $m_t^2$ which means
that ${\cal O}(\epsilon)$ contributions are only needed for the
${\cal O}(s/m_t^2)$ term, which are much simpler to compute than the
${\cal O}(m_t^0)$ terms.  Note that our explicit expressions for
$\vec{I}_{33}^{\,(d)}$ contain constants and functions which have at most
transcendental weight four. For details on their computation we refer to
Ref.~\cite{Mishima:2018}.

For the three $5\times5$ blocks there are similar transformation as
in~(\ref{eq:basis-change33}) and the same procedure is performed as described
above.


\end{appendix}


\end{document}